\documentclass[showpacs,amsmath,amssymb,twocolumn,floatfix,prl]{revtex4}
\usepackage[dvips]{graphicx}
\input{epsf}

\begin{document}

\title{Quantum computation and analysis of Wigner and Husimi functions: \\
toward a quantum image treatment}


\author{M. Terraneo, B.Georgeot and D. L. Shepelyansky}
\affiliation{Laboratoire de Physique Th\'eorique, UMR 5152 du CNRS, 
Universit\'e Paul Sabatier, 31062 Toulouse Cedex 4, France}
\homepage[]{http://www.quantware.ups-tlse.fr}

\date{December 15, 2004}

\begin{abstract}
We study the efficiency of quantum algorithms
which aim at obtaining phase space distribution functions of
quantum systems.  Wigner and Husimi functions are considered.
Different quantum algorithms are envisioned to build these functions, and
compared with the classical computation.  Different procedures to extract
more efficiently information from the final wave function of these algorithms
are studied, including
coarse-grained measurements, amplitude amplification and measure
of wavelet-transformed wave function.
The algorithms are analyzed and numerically tested on a complex
quantum system showing different behavior depending on parameters, namely
the kicked rotator.
The results for the Wigner function show in particular
that the use of the quantum
wavelet transform gives a polynomial gain over classical computation.
For the Husimi distribution, the gain is much larger than for the 
Wigner function, and is bigger with the help of amplitude amplification
and wavelet transforms. We also apply the same set of techniques to 
the analysis of real images. The results show that the use of 
the quantum wavelet transform allows to lower dramatically
the number of measurements needed, but at the cost of a large loss of 
information.

\end{abstract}
\maketitle

\section{I.Introduction}

In the recent years, the study
of quantum information \cite{nielsen} has attracted more and more interest.  
In this field,
quantum mechanics is used 
to treat and manipulate information.  Important applications
are quantum cryptography, quantum teleportation and quantum computation.
The latter takes advantage of the laws of quantum mechanics 
to perform computational tasks
sometimes much faster than classical devices.  A famous example is provided
by the problem of factoring large integers, useful for public-key cryptography,
which can be solved with exponential efficiency by Shor's algorithm 
\cite{shor}.  Another example is the search of an unstructured list, which
was shown by Grover \cite{grover} to be quadratically faster on quantum 
devices.  In parallel, investigations of the simulation of
quantum systems on quantum computers showed that the evolution of
a complex wave function can be simulated efficiently for an exponentially 
large Hilbert space with polynomial resources
\cite{lloyd,schack,GS,song,complex,pomeransky}.  
Still, 
there are many open questions which remain unanswered.  In particular, 
it is not always clear how to perform an
efficient extraction of information from such a complex quantum mechanical 
wave function once it has been
evolved on a quantum computer.  More generally, the same problem appears
for quantum algorithms manipulating large amount of classical data.

In the present paper, we study 
different algorithmic processes which perform this task.  We focus on
the phase space distribution (Wigner and Husimi functions) \cite{wigner,husimi}
These functions provide
a two-dimensional picture of a one-dimensional wave function, and can be
compared directly with classical phase space distributions.  
They have also 
been shown in \cite{Levi,harper} to be stable with respect to various
quantum computer error models.
Different phase space representation 
which can be implemented efficiently on a quantum computer will be explored,
 first the discrete
Wigner transform, for which an original algorithm will be presented,
and then a Husimi-like transform, first introduced in this context in 
\cite{frahm}.  
Recent proposals \cite{frahm,pazwigner,saraceno}
gave methods to measure or construct 
Wigner and Husimi functions on a quantum computer, using for
example phase space tomography.
These method will be analyzed and compared with new strategies, in order to
identify the most efficient algorithms.
Different techniques will be tested in order to extract 
information, namely measure of an ancilla qubit, measurement
of all qubits, coarse grained measurement,
and the use of  amplitude amplification \cite{amplification}.
In addition, we will analyze the use of the wavelet transform
to compress information and minimize the number of measurements.
Indeed, wavelet transforms \cite{Daub,meyer}
are used in a large number of applications
involving classical data treatment, in particular they allow
to reach large compression rates for classical images 
in standards like MPEG.
Quantum wavelet transforms have been built and implemented
\cite{WT1,WT2,WT3,terraneo}, and it was shown that they can
be applied on an exponentially large vector in a 
polynomial number of operations.  
Numerical computations will enable us to quantify the efficiency of
each method for a specific complex quantum system, namely the kicked
rotator.  In general, it will be shown that a polynomial gain can be reached
with several strategies.
Since a quantum phase space distribution 
can be considered as an example of a two-dimensional picture,
we discuss in a subsequent section the use of the same techniques
to treat images encoded on the wave function of a quantum computer,
in a way similar to what is done in classical image analysis. This for
example could be applied to images transmitted through quantum imaging
\cite{kolobov}.

\section{II. Quantum phase space distributions for a chaotic quantum map}

Classical Hamiltonian mechanics is built in phase space, dynamics
being governed by Hamilton's equation of motion.  Classical
motion can be described through the evolution of phase space (Liouville)
distributions.  On the other hand,
phase space is a peculiar notion in quantum mechanics since
$p$ and $q$ do not commute.  A wave function is naturally described
in a Hilbert space, for example position alone or momentum alone.
Nevertheless, it has been known since a long time that it is
 possible to define
functions of $p$ and $q$ which can be thought 
as quantum phase space distributions.  The most commonly used is the
Wigner function \cite{wigner}, defined for the wave function 
$\psi$ of a continuous system by: 

\begin{equation}
\label{wigner}
W(p,q)= \int
\frac{e^{-\frac{i}{\hbar}p.q'}}{\sqrt{2\pi\hbar}}
\psi(q+\frac{q'}{2})^{*}\psi(q-\frac{q'}{2}) dq'
\end{equation}
 
This function involves the two variables position $q$ and momentum $p$
in a symmetric way (although it is not immediately apparent in the formula
(\ref{wigner})), and shares
some properties with classical phase space probability distributions.
Indeed, it is a {\em real} function, and satisfies $\int W(p,q) dq=|\psi(p)|^2$
and $\int W(p,q) dp=|\psi(q)|^2$.  However, it cannot be identified
with a probability distribution since it can take {\em negative} values.
The Wigner function has been measured experimentally in atomic systems, and
such negative values have been reported \cite{negative}.   

Although the Wigner function can take negative values, it can be shown
that coarse graining this function over cells of size
$\hbar$ always leads to nonnegative values.  Therefore a smoothing
of (\ref{wigner}) by appropriate functions will lead to a function 
of $p$ and $q$ with no negative values.  An example of such a function is
given by the {\em Husimi distribution} (see e.g. \cite{husimi}) which
uses a Gaussian smoothing.  A further example using another smoothing function
was discussed in \cite{frahm}.

In the following sections, we will study the evaluation of
 such quantum phase space distributions of wave functions on a quantum
computer.  This will be performed using a specific example, 
namely the kicked rotator model.  
This system corresponds to the quantization of 
the Chirikov standard map \cite{qchaos,lichtenberg}
$\bar{n} = n + k \sin{  \theta }; \;\;\; \bar{\theta} = \theta + T \bar{n}$
where $(n,\theta)$ are the conjugated (action-angle) variables.

\begin{figure}[h!]
\includegraphics[width=.49\linewidth]{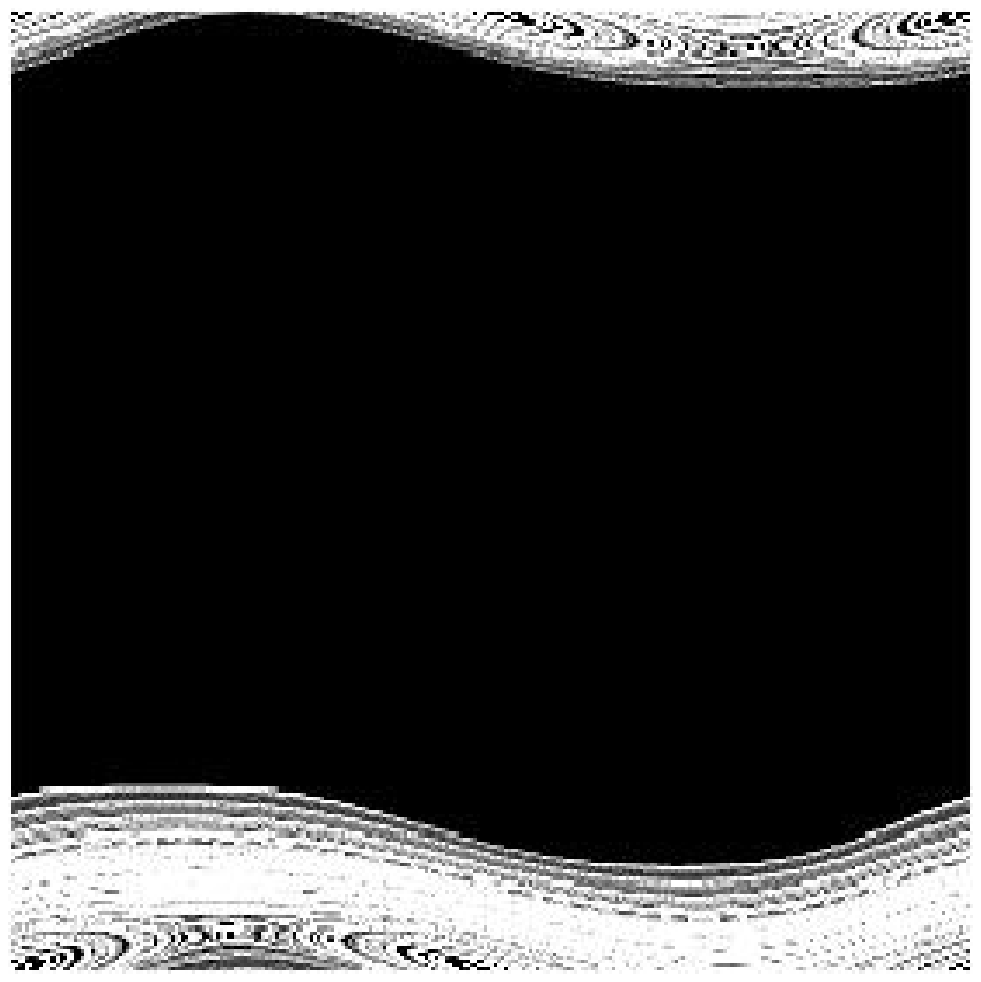}
\hfill
\includegraphics[width=.49\linewidth]{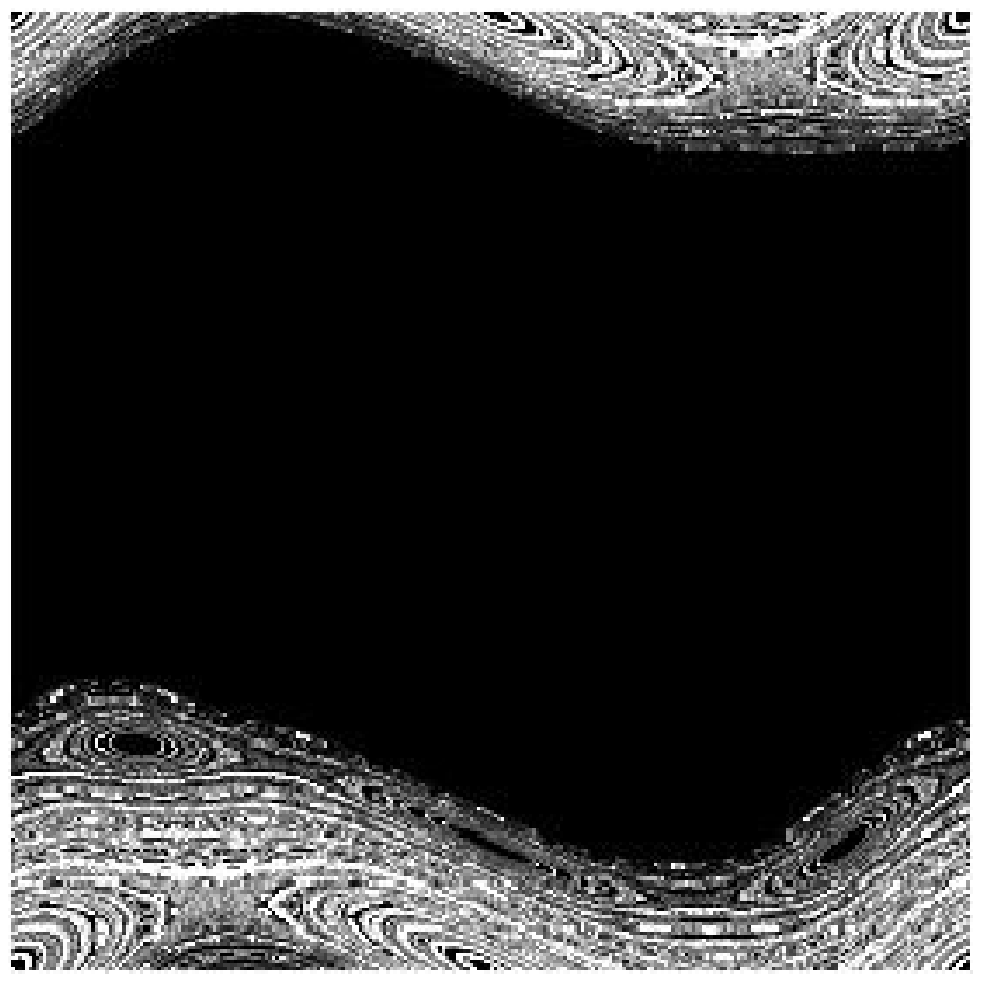}
\includegraphics[width=.49\linewidth]{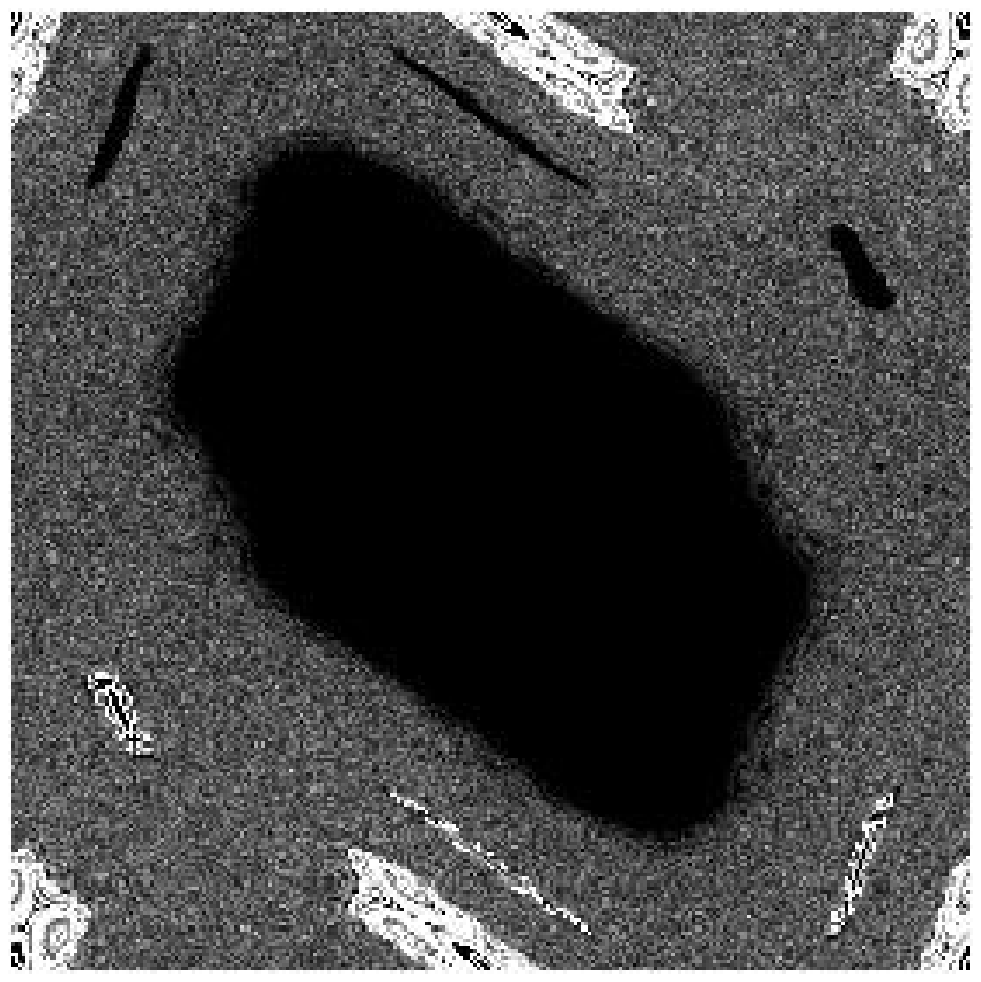}
\hfill
\includegraphics[width=.49\linewidth]{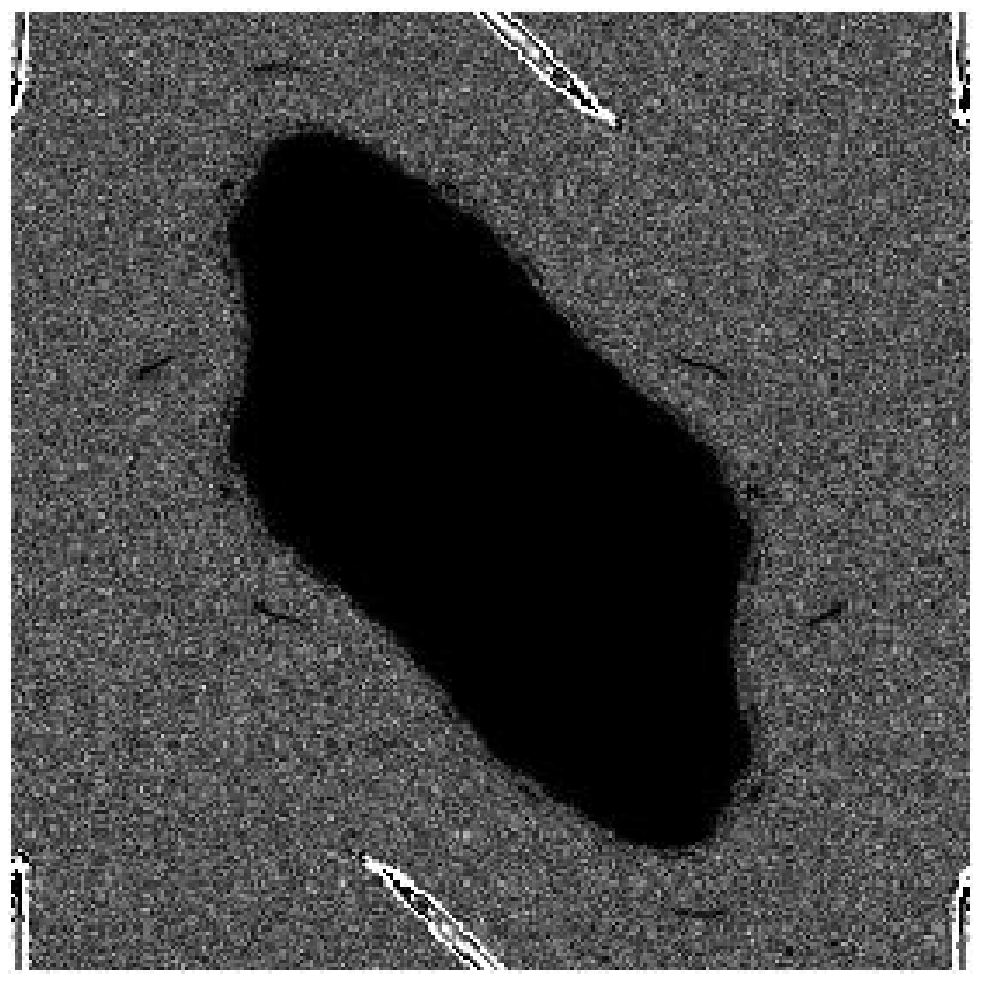}
\caption{Classical phase space distribution for the standard map with
$K=0.5$ (top left), $K=0.9$ (top right), $K=1.5$ (bottom left),
$K=2$ (bottom right). Black is zero probability, 
white is maximal probability. As initial state we chose a uniform distribution
on the set $-\pi \le p \le -3/4\pi$, $ 0 \le x \le 2 \pi$, and 
$1000$ iterations of the standard map were performed.
}
\label{classical}
\end{figure}

The classical standard map depends only on the parameter $K=kT$.
The system undergoes a transition from integrability ($K=0$)
to more and more developed chaos when $K$ increases, following the
Kolmogorov-Arnold-Moser theorem.  Chaotic zones get larger and larger
until the value $K=K_g \approx 0.9716...$ is reached, where global chaos 
sets in, but a complex hierarchical structure of
integrable islands surrounded by chaotic layers is still present.  For 
$K \gg K_g$, the chaotic part covers most of the phase space.
This system has been used for example as a model
of  particle confinement in magnetic traps, beam dynamics in
accelerators or comet trajectories  \cite{lichtenberg}. Its phase space is
a cylinder (periodicity in $\theta$), and since the map
is periodic in $n$ with period $2\pi/T$, 
phase space structures repeat themselves in the $n$
direction on each cell of size $2\pi/T$. Fig.\ref{classical}
shows one such phase space cell for various values of the
parameter $K$, showing the different regimes from quasi-integrability
(many invariant curves preventing transport in the momentum direction)
 to a mixed regime with a large chaotic domain.  

The quantum version of the standard map \cite{qchaos}
gives a unitary operator acting 
on the wave function $\psi$ through:
\begin{eqnarray} 
\label{qmap}
\bar{\psi} = \hat{U} \psi =  e^{-ik\cos{\hat{\theta}}}
 e^{-iT\hat{n}^2/2} \psi,
\end{eqnarray}
where $\hat{n}=-i \partial / \partial \theta $, $\hbar=1$,
and $\psi(\theta+2\pi)=\psi(\theta)$.

The quantum dynamics (\ref{qmap}) depends on the
two parameters $k$ and $T$, $T$ playing the role of an effective $\hbar$.
The classical limit is $k \rightarrow \infty$, $T \rightarrow 0$ while
keeping $K=kT=$ constant.  

This quantum kicked rotator (\ref{qmap}) is described
by quite simple equations, making it practical for
numerical simulations and quantum computing.  Nevertheless,
it displays a wealth of different behaviors depending on the values
of the parameters.  Indeed, classical dynamics
 undergoes a transition
from integrability to fully developed chaos
with intermediate mixed phases between these two regimes.
Wave functions show complex structures related to the classical phase
space corresponding to these different cases. In addition, for large $K$
where classical dynamics is strongly chaotic, quantum interference
can lead to exponential localization of wave functions.  This phenomenon is
related to  the Anderson localization of electrons in solids, and therefore
enables to study this important solid state problem, which is still
the subject of active research.  The kicked rotator can also model
the microwave
ionization of Rydberg atoms \cite{IEEE}, and has been 
experimentally realized with cold atoms \cite{raizen}.  For all these reasons,
it has been the subject of many studies, and can be considered
as a paradigmatic model of quantum chaos. 

In \cite{GS,Levi} it was shown
that evolving a $N$-dimensional wave function through the map
(\ref{qmap}) can be done with only $O(\log N)$ qubits
and $O((\log N)^3)$ operations on a quantum computer
(compare with $O(N\log N)$ operations
for the same 
simulation on a classical computer).  Another quantum algorithm developed
in \cite{pomeransky} enables to perform the same quantum evolution (albeit
approximately)
with $O((\log N)^2)$ operations.
This system can therefore be simulated efficiently on a quantum computer,
and can be used as a good test ground for assessing the complexity
of various quantum algorithms for quantum phase space distributions.

In the following sections, we will study the efficiency of 
various quantum algorithms to obtain various information about
the quantum phase space distribution functions.  
The simulation of a quantum system on a quantum computer
based on qubits implies that the system is effectively
discrete and finite.  We therefore close the phase space in the momentum
direction through periodic boundary conditions.  We will concentrate
on the regime where $T=2\pi/N$, $N$ being the Hilbert space dimension.
This implies that the 
phase space contains only one classical cell, and increasing
the number of qubits at $K$ constant 
decreases the effective $\hbar$ keeping the
classical dynamics constant.  Different $K$ values enable to probe 
various dynamical regimes, from integrability to chaos. 
The localization length in this regime
becomes quickly larger than the system size for small number
of qubits, thus
allowing to explore the complexity of a chaotic wave function.  Indeed, 
in the localized regime, the most important information resides not so much
in such distributions, but in the localization properties, and their
measurement on a quantum computer was already 
analyzed in \cite{loclength, harper}.

\begin{figure}[h!]
\includegraphics[width=.49\linewidth]{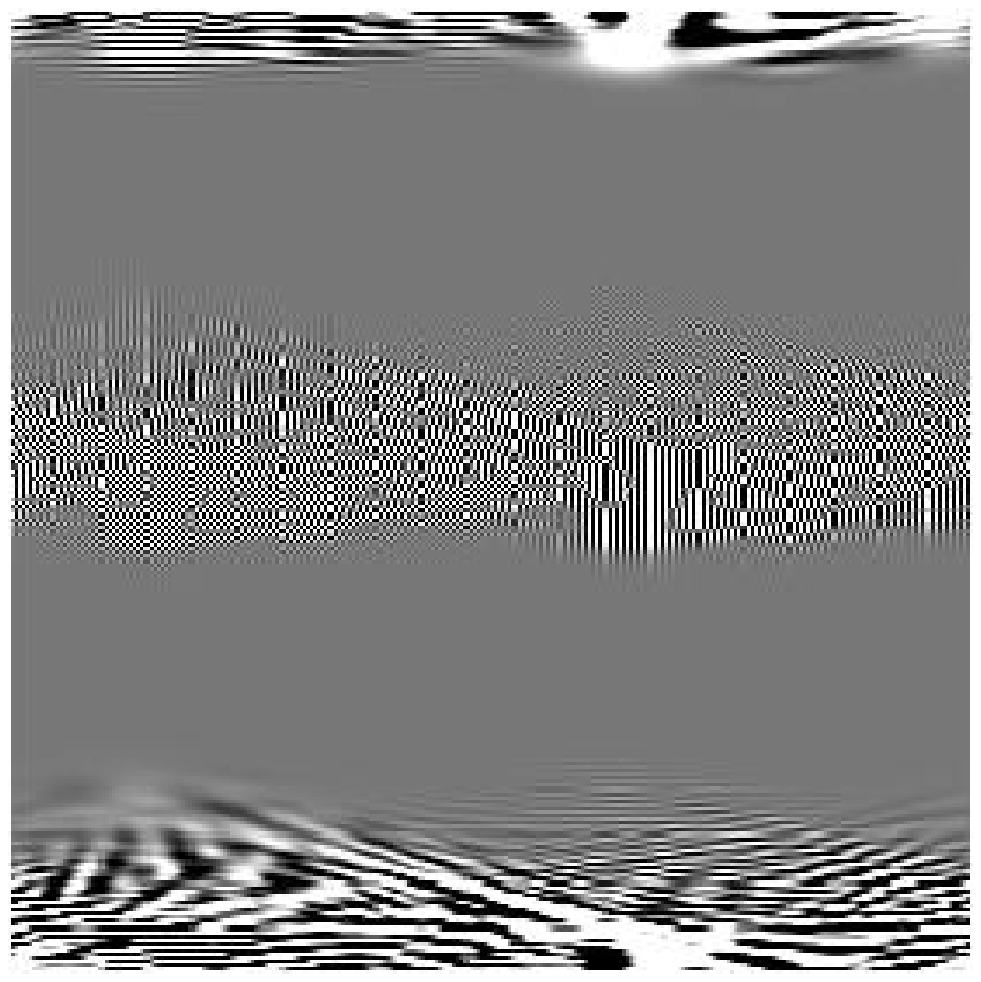}
\hfill
\includegraphics[width=.49\linewidth]{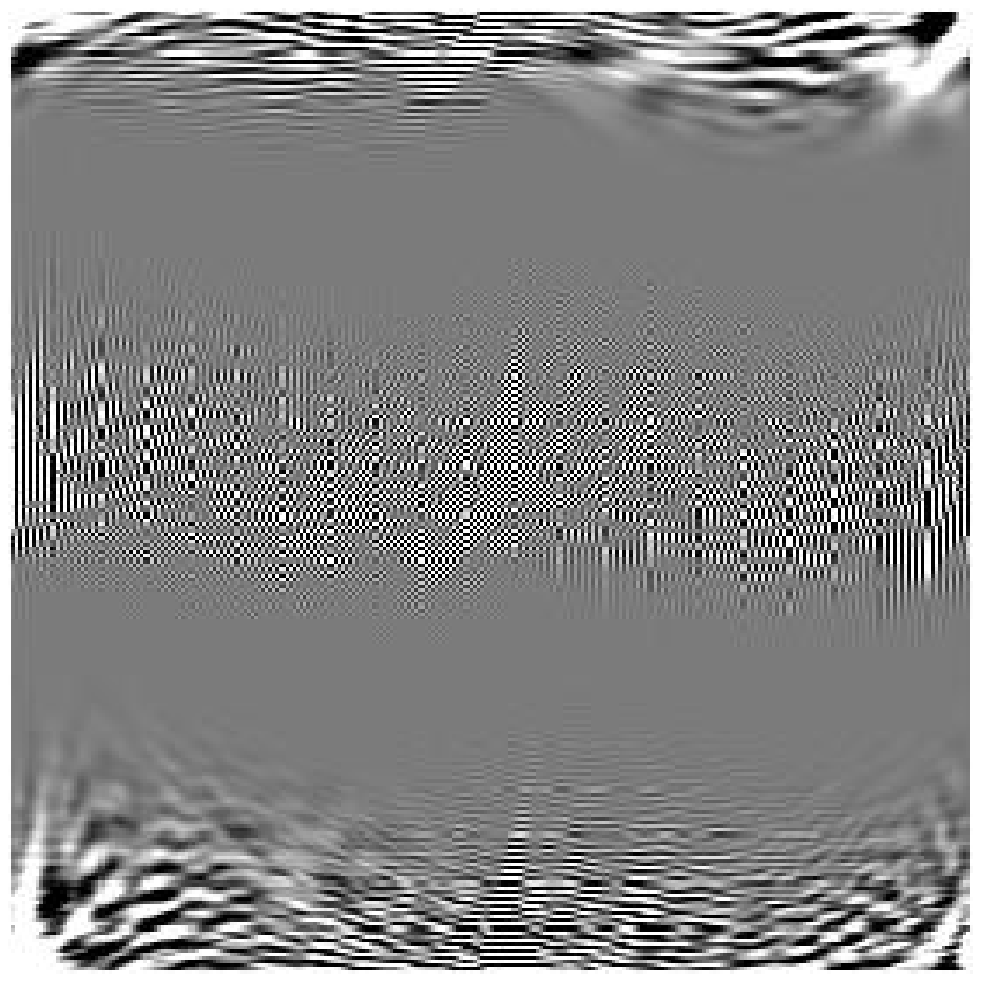}
\includegraphics[width=.49\linewidth]{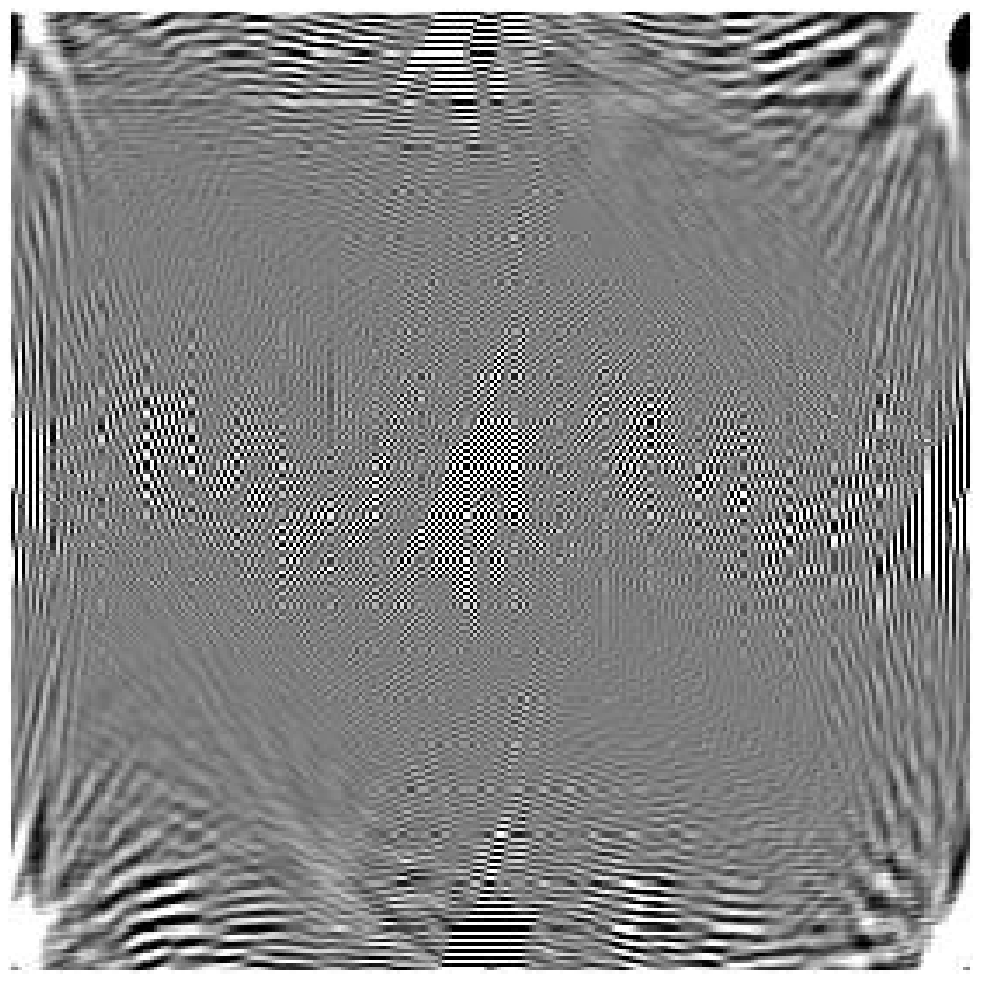}
\hfill
\includegraphics[width=.49\linewidth]{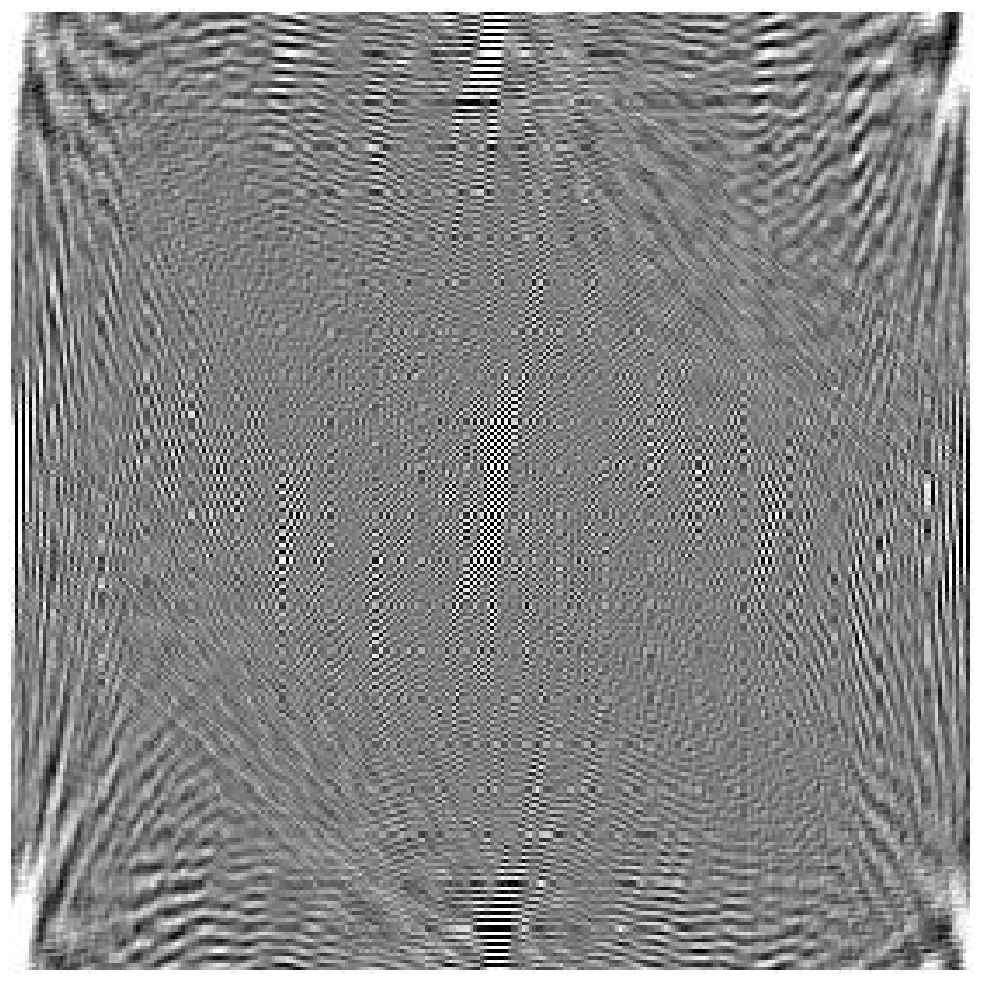}
\caption{
Wigner function for the quantum kicked rotator with parameters of 
Fig.\ref{classical} 
$K=0.5$ (top left), $K=0.9$ (top right), $K=1.5$ (bottom left),
$K=2$ (bottom right). Here $T=2\pi/N$, where $N=2^{n_q}$, with $n_q=7$.
The whole Wigner function (on a $2 N \times 2 N$ lattice) is plotted. White 
marks positive maximal values, black negative values. Initial state 
is uniformly spread
on the set $0 \le n < N/8$
(corresponding to the initial classical distribution in Fig.1)
(this state can be built efficiently from $|n=0\rangle$
by $n_q -3$ Hadamard gates), and Wigner 
function is computed after $1000$ iterations of (\ref{qmap}).
}
\label{wignerf}
\end{figure}

For such a quantum system on a $N$ dimensional Hilbert space,
the general formalism of Wigner functions should be adapted.
In particular, it is known that it should be constructed on
$2N \times 2N$ points (see e.g. \cite{discrete}).  
For the kicked rotator, the formula for the discrete Wigner function is:
\begin{equation}
\label{wigner2}
W(\Theta,n)= 
\sum_{m=0}^{N-1} \frac{e^{-\frac{2i\pi}{N}n(m-\Theta/2)}}{2N}
\psi(\Theta-m)^{*}\psi(m),
\end{equation}
with $\Theta=\frac{N\theta}{2\pi}$.  

The Wigner function provides a pictorial representation of a wave function
which can be compared with the classical phase space distribution,
(see example in Fig.\ref{wignerf}), although quantum oscillations are present.

\section{III. Measuring the Wigner distribution}

In \cite{pazwigner} the first quantum
 algorithm was set up which enables to measure
the value of the Wigner function at a given phase space point.  
The algorithm adds one ancilla qubit to the system and proceeds by 
applying one Hadamard gate to the ancilla qubit, then a certain operator
$U(\Theta,n)$ is applied to the system controlled by the value of the ancilla
qubit.  After a last Hadamard gate is applied to the ancilla, its
expectation value is $<\sigma^z>=Re[Tr(U(\Theta,n)\rho)]=2NW(\Theta,n)$
where $\rho$ is the density matrix and $N=2^{n_q}$ is the dimension
of the Hilbert space.
One iteration of this process requires only a logarithmic number of gates.
Nevertheless, the total complexity of the algorithm 
may be much larger,
since measuring $<\sigma^z>$ may require a very large number of measurements.
This can be probed only through careful estimation of the asymptotic
behavior of individual values of the Wigner function.

A drawback of the approach of \cite{pazwigner} is that it does not allow
easily further treatment on the Wigner function which may improve
the total complexity of the algorithm.  To this aim, the simplest way
is to build explicitly the Wigner 
transform of the wave function as amplitudes of a register.  This enables to
use additional tools (amplitude amplification, wavelet transforms)
which may increase the
speedup over classical computation, as we will see.  

Such an explicit construction of the Wigner function directly on the
registers of the quantum computer is indeed possible in the following way.
To get the Wigner function of $\hat{U}^t |\psi_0\rangle$ ($t$ iterations
of an original wave function $|\psi_0\rangle$ through (\ref{qmap})),
we start from an initial state (for example in $n$
representation)
$|\psi_0\rangle \otimes |\psi^*_0 \rangle$
$=\sum_{i=0}^{N-1} a_i |n_i\rangle \otimes 
\sum_{j=0}^{N-1} a^*_j |n_j\rangle$ $=\sum_{i=0}^{N-1}\sum_{j=0}^{N-1} 
a_i a^*_j|n_i\rangle  |n_j\rangle$.  This needs $2n_q$ qubits
to hold the values of the wave function on a $N$-dimensional Hilbert space,
where $N=2^{n_q}$.
Then we apply the algorithm implementing the
kicked rotator evolution operator $\hat{U}$ developed in \cite{GS}
to each subsystem independently.
This operator can be described as multiplication by phases followed by 
a quantum Fourier transform (QFT).
The multiplication by phases of each coefficient keeps the factorized 
structure. The QFT mixes only states with the same value of the other register
attached, and therefore also keeps the factorized form.  
Let us see how it works for one iteration:

\begin{equation}
\sum_{i=0}^{N-1}\sum_{j=0}^{N-1} 
a_i a^*_j|n_i\rangle  |n_j\rangle
\rightarrow \sum_{i=0}^{N-1}\sum_{j=0}^{N-1} 
e^{-iTn_i^2/2}a_i a^*_j|n_i\rangle  |n_j\rangle
\nonumber
\end{equation}

(multiplication by $e^{-iT\hat{n_i}^2/2}$)
\begin{equation}
=\sum_{j=0}^{N-1}(\sum_{i=0}^{N-1} e^{-iTn_i^2/2}a_i |n_i\rangle)
a^*_j|n_j\rangle
\rightarrow \sum_{j=0}^{N-1} (\sum_{i=0}^{N-1}
b_i |\theta_i\rangle) a^*_j |n_j\rangle
\nonumber
\end{equation}
(QFT with respect to $n_i$)
\begin{equation}
\rightarrow \sum_{j=0}^{N-1}(\sum_{i=0}^{N-1}
e^{-ik\cos{\theta_i}}b_i |\theta_i\rangle ) a^*_j|n_j\rangle
\nonumber
\end{equation}
(multiplication by $e^{-ik\cos{\theta_i}}$)
\begin{equation}
\rightarrow \sum_{j=0}^{N-1}(\sum_{i=0}^{N-1}
c_i |n_i\rangle)a^*_j  |n_j\rangle
\nonumber
\end{equation}
(QFT with respect to $\theta_i$)
\begin{equation}
=(\sum_{i=0}^{N-1}
c_i |n_i\rangle)\otimes (\sum_{j=0}^{N-1}  a^*_j  |n_j\rangle)
=\hat{U}|\psi_0\rangle \otimes |\psi^*_0 \rangle
\nonumber
\end{equation}
We can thus get 
$\hat{U}^t |\psi_0\rangle \otimes \hat{U}^{*t} |\psi^*_0 \rangle$
by applying the process several times.  
This can be done in a number of gates polynomial in $n_q$
($O(tn_q^3)$ if we use the algorithm of \cite{GS} for implementing $\hat{U}$).

From such a state it is possible to build efficiently the state
$\sum_{\theta,n} W(\theta,n) |\theta\rangle |n\rangle$.  Indeed,
building the Wigner transform can be done through a partial Fourier transform.
To see this, we start from the state in $\theta$ representation, i.e. 
$|\psi\rangle \otimes |\psi^* \rangle$=
$\sum_{\theta,\theta'} \psi(\theta)\psi^* (\theta')
|\theta \rangle |\theta'\rangle$.  Then we add an extra
qubit to the first register (needed to get values
of $\theta +\theta'$ between $0$ and $2N-1$) and realize the transformation:

$\sum_{\theta,\theta'} \psi(\theta)\psi^* (\theta')
|\theta \rangle |\theta'\rangle$ 
$\rightarrow 
\sum_{\theta,\theta'} \psi(\theta)\psi^* (\theta')
|\theta +\theta'\rangle |\theta'\rangle$ (addition)

Let us call $\Theta=\theta+\theta'$; then the state can be written
$\sum_{\Theta,\theta'} \psi(\Theta-\theta')\psi^* (\theta')
|\Theta\rangle |\theta'\rangle$
Then we realize a QFT of the second register only.  The result is:
$\sum_{\Theta} \sum_{n} (\sum_{\theta'} 
e^{-\frac{2i\pi}{N}n\theta'} \psi(\Theta-\theta')\psi^* (\theta'))
|\Theta\rangle |n\rangle$
=$2\sqrt{N}\sum_{\Theta} \sum_{n} W(\Theta,n)e^{-\frac{2i\pi}{N}n\Theta/2}
|\Theta\rangle |n\rangle$
where $\Theta$ varies from $0$ to $2N-1$ and $n$ from $0$ to $N-1$.
To get the Wigner function on a $2N \times 2N$ grid, we need first
to add an extra qubit in the state $|0\rangle$ and apply a Hadamard
gate to it.  If we interpret it as the most significant digit of $n$,
the resulting state is 
$\sqrt{2N}
\sum_{\Theta} \sum_{n=0}^{N-1} W(\Theta,n)e^{-\frac{2i\pi}{N}n\Theta/2}
|\Theta\rangle |n\rangle +\sqrt{2N}
\sum_{\Theta} \sum_{n=N}^{2N-1} W(\Theta,n)e^{-\frac{2i\pi}{N}(n-N)\Theta/2}
|\Theta\rangle |n\rangle$.
The final step consists in multiplying by the phases 
$e^{-\frac{2i\pi}{N}n\Theta/2}$ and $e^{-\frac{2i\pi}{N}(n-N)\Theta/2}$,
which can be made by $n_q^2$ application of two-qubit gates 
(controlled phase-shifts).
The final state is 

\begin{equation}
\label{wignerimage}
|\psi_f \rangle=\sqrt{2N}\sum_{\Theta=0}^{2N-1} \sum_{n=0}^{2N-1} W(\Theta,n)
|\Theta\rangle |n\rangle
\end{equation}

One can check that the normalization is correct
since it is known in general that 
$\sum_{\Theta=0}^{2N-1} \sum_{n=0}^{2N-1} W(\Theta,n)^2=1/2N$.

\begin{figure}[h]
\includegraphics[width=.8\linewidth]{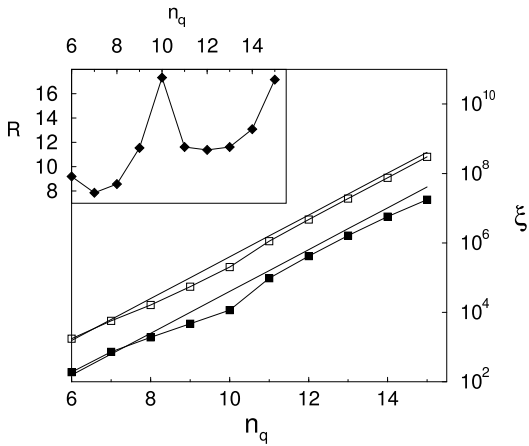}
\caption{Main plot: scaling of the IPR $\xi$ vs. $n_q$ 
for the Wigner function (empty 
squares) and for the wavelet transform of the Wigner function (full squares).
The full straight lines represent the law $N^2$, $N=2^{n_q}$. Here $K=0.5$.
In the inset, the ratio $R$ between IPR of Wigner function and 
wavelet transformed Wigner function is plotted. Parameters, 
number of iterations and initial state are the same as in Fig.\ref{wignerf}.
}
\label{IPR05}
\end{figure}

\begin{figure}[h]
\includegraphics[width=.8\linewidth]{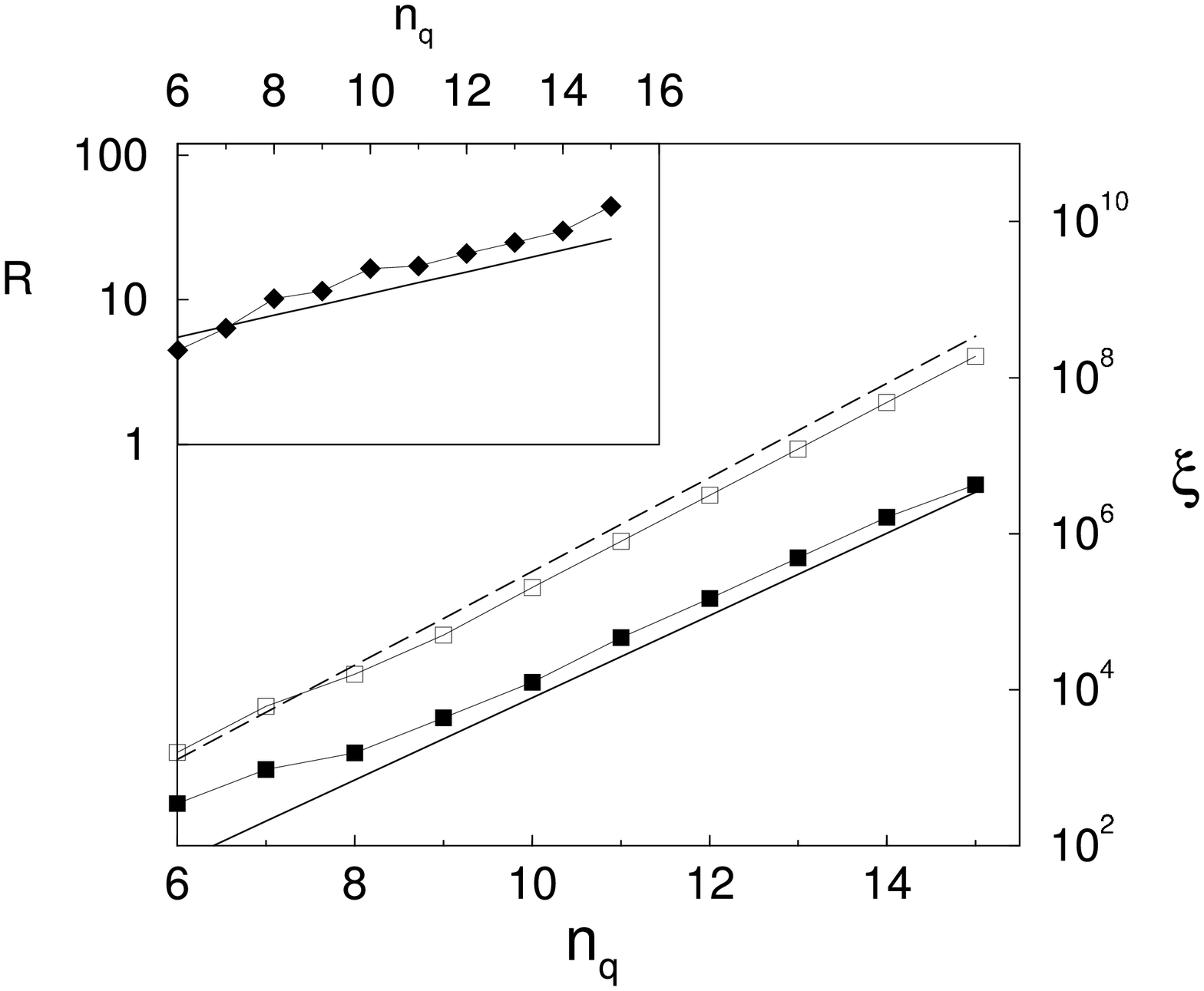}
\caption{Main plot: scaling of the IPR $\xi$ vs. $n_q$ 
for the Wigner function (empty 
squares) and for the wavelet transform of the Wigner function (full squares).
The full straight line represents the law $N^{1.75}$, while the dashed line
represents $N^{2}$, $N=2^{n_q}$. Here $K=0.9$.
In the inset, the ratio $R$ between IPR of Wigner function and 
wavelet transformed Wigner function is plotted. 
The full line represents the scaling $N^{0.25}$. Parameters, 
number of iterations and initial state are the same as in Fig.\ref{wignerf}.
}
\label{IPR09}
\end{figure}

\begin{figure}[h]
\includegraphics[width=.8\linewidth]{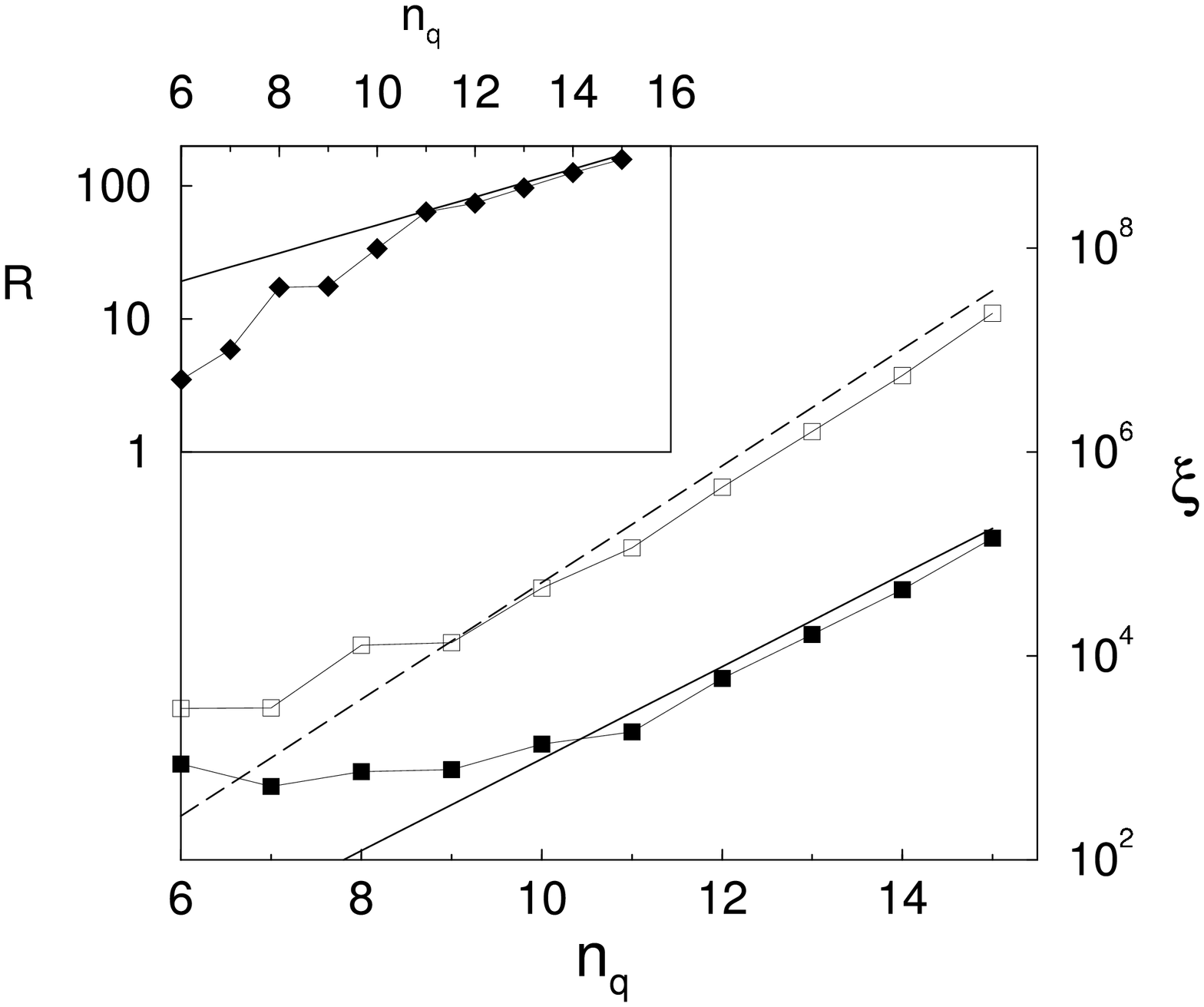}
\caption{Main plot: scaling of the IPR $\xi$ vs. $n_q$ 
for the Wigner function (empty 
squares) and for the wavelet transform of the Wigner function (full squares).
The full straight line represents the law $N^{1.5}$, while the dashed line
represents $N^{1.9}$, $N=2^{n_q}$. Here $K=1.5$.
In the inset, the ratio $R$ between IPR of Wigner function and 
wavelet transformed Wigner function is plotted.
The full line represents the scaling $N^{0.4}$. Parameters, 
number of iterations and initial state are the same as in Fig.\ref{wignerf}.
}
\label{IPR15}
\end{figure}

\begin{figure}[h]
\includegraphics[width=.8\linewidth]{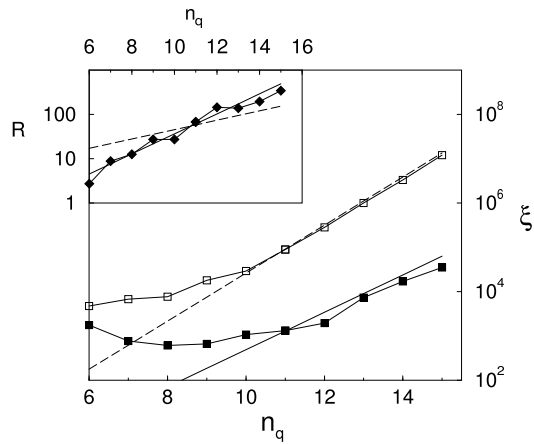}
\caption{Main plot: scaling of the IPR $\xi$ vs. $n_q$ 
for the Wigner function (empty 
squares) and for the wavelet transform of the Wigner function (full squares).
The full straight line represents the law $N^{1.4}$, while the dashed line
represents $N^{1.8}$, $N=2^{n_q}$. Here $K=2$.
In the inset, the ratio $R$ between IPR of Wigner function and 
wavelet transformed Wigner function is plotted.
The full line represents the scaling $N^{0.75}$,while the dashed line 
represents  $N^{0.35}$. Parameters, 
number of iterations and initial state are the same as in Fig.\ref{wignerf}.
}
\label{IPR2}
\end{figure}

The advantage of this procedure in comparison to the one in 
\cite{pazwigner} resides in the fact that individual values
of the Wigner function are now encoded in the components of the wave function.
This is in general a natural way to encode an image on a wave function:
each basis vector corresponding to a position in phase space is associated
with a coefficient giving the amplitude at this location.  This way
of encoding the Wigner function enables to perform some further operations
to extract information efficiently through quantum measurements.  We will
envision three different strategies: direct measurements
of each qubit, amplitude amplification and wavelet transform.
The data of Fig.\ref{IPR05}-\ref{IPR2}
 will enable us to compare these different strategies
for different physical regimes of the kicked rotator model,
with various levels of chaoticity.
The quantity plotted is the inverse participation ratio (IPR).  
For a wave
function $|\psi\rangle= \sum_{i=1}^{N} \psi_i |i\rangle$, where 
$|i\rangle$ is some basis,
the inverse participation ratio is
$\sum |\psi_i|^2/(\sum |\psi_i|^4)$ and measures the number of
significant components in the basis $|i\rangle$.  The Wigner function verifies
the sum rules $\sum W_i=1$ and $\sum W_i^2=\frac{1}{N}$.
Following \cite{Levi} we are lead by analogy to define the
inverse participation ratio for the Wigner function,
by the formula $\xi=1/(N^2\sum W_i^4)$.
If the Wigner function is composed of $N$ components of equal weights
$1/N$, then $\xi=N$,
whereas $N^2$ components of equal weights (in absolute value)
$1/N^{3/2}$
give $\xi=N^2$.  Thus the IPR $\xi$ gives an estimate of
the number of the main components of the Wigner function.


To compare classical and quantum computation of this problem, we first 
should assess the complexity of obtaining the Wigner function on a 
classical computer.  For a $N$-dimensional wave function, iterating
$t$ times the map (\ref{qmap}) costs $O(tN\log N )$ operations.
Then getting all values of $W$ needs to perform $N$ Fourier transforms,
requiring $O(N^2 \log N )$ operations. The same is true for obtaining
the largest values of $W$, if one does not know where they are: only
the computation of all of them and subsequent sorting can provide them.
Thus in both cases classical complexity is of the order $O(N^2 \log N )$.
This asymptotic law changes if one is interested in a {\em single} value
of the Wigner function at some predetermined $(\Theta,n)$ value.  In this case,
only one Fourier transform is actually needed, so the classical complexity
becomes of order $O(N \log N )$.

As concerns the quantum computer, we have to clarify the measurement 
protocol to assess the complexity of the algorithm.
The most obvious strategy consists in measuring all the qubits after
explicit construction of the wave function (\ref{wignerimage}) and
accumulating statistics until a good precision is attained on all 
values of the Wigner function.   From Fig.\ref{IPR05}-\ref{IPR2}
(empty squares),
we can see that in the four physical regimes considered, the IPR
scales approximately as $N^2$.  This implies that the Wigner function is
spread out on the $N^2$ components, each term having comparable
amplitude $W_i\sim N^{-3/2}$.  This needs $N^2$ measurements
to get a good precision.  The number of quantum operations is therefore
$O(tN^2)$ ($N^2$ repetitions of $t$ iterations) up to logarithmic factors.
This should be compared with the classical complexity of obtaining all
values of the Wigner function, or only the largest ones, which both 
are of order $O(N^2\log N)$.
This makes the quantum method no better than the classical one, 
albeit the quantum computer needs a logarithmic number of qubits 
whereas the classical computer needs exponentially more bits ($N$ bits
versus $\log N$ qubits).  This can translate in an
improvement in effective computational time by for example
distributing the computation over subsystems of
qubits, and making simultaneous 
measurements, but this is obviously quite cumbersome.

Still, it can be remarked that for the values of $K$ for which
the system is most chaotic, the IPR scales with a 
slightly lower power $N^{\alpha}$ with $\alpha\approx 1.8-1.9$.  
If this is verified asymptotically,
then the quantum algorithm need only $O(tN^{\alpha})$ operations,
and a small gain of $N^{2-\alpha}$ is realized.  It is interesting to note
that if the classical algorithm to compute the evolution of the map
were more complex, i.e. of order $O(tN^k)$ with $k>2$, then in this 
case the quantum algorithm will be better by an additional
factor of $N^{k-2}$.

The phase space tomography method of \cite{pazwigner} requires to measure 
$<\sigma^z>$ of an ancilla qubit, with $<\sigma^z>=NW(\Theta,n)$.  Thus
$<\sigma^z> \sim N^{-1/2}$, a value which requires $N$ measurements 
to be reasonably assessed.  This should be compared with the classical cost
of obtaining the value of the Wigner function at a predetermined location,
which is of order $O(N \log N )$.  Again, the method is not better than the
classical one, although it uses a logarithmic number of qubits which
can translate into an 
improvement in effective computational time by
distributing the quantum computation.  Similarly, when the IPR
scales as $N^{\alpha}$ with $\alpha < 2$, then the quantum algorithm is better
by a factor of $N^{2-\alpha}$.  For other maps for which
classical simulation is of order $O(tN^k)$ with $k>2$, the quantum algorithm 
will be better by an additional factor of $N^{k-2}$.

It is possible to use coarse-grained
measurements in order to decrease the number
of measurements of the wave function ({\ref{wignerimage}).  
To this aim, one measures
only the first $n_f$ qubits with $n_f<n_q$ ($N=2^{n_q}$).  
This gives the
integrated probability inside the $2^{2n_f}$ cells (sum of $2^{2n_q -2n_f}$
probabilities $|W(\Theta,n)|^2$)) in a number of measurements which scales 
with the number of cells and not any more with the number of qubits.
This is possible if the wave function of the computer encodes
the full Wigner function in its components, as in the algorithm exposed above.
In principle, the complexity is $O(2^{2n_f})$ and a gain compared
to classical computation can be obtained.
There is a possibility of exponential gain with this strategy,
since by fixing $n_f$ and letting $n_q$ increase, measuring the integrated
probability becomes polynomial in $n_q$.  Still, the precision is also
polynomial, and it is possible that semiclassical methods enable to 
get such approximate quantities since with $n_q \rightarrow \infty$ the 
value of $\hbar$ becomes smaller and smaller and the system is
well approximated by semiclassical calculations.  If this holds, the
advantage of quantum computation may be less spectacular.

A similar method can be applied to the phase space tomography method of
\cite{pazwigner}, but with a different result.  
In \cite{saraceno} it is explained that one can
compute averages of Wigner function on a given rectangular area
by using an ancilla qubit.
The process gives $<\sigma^z> = 2N\sum W(\Theta,n)/N_P$, where $N_P$ 
is the number of points over which the summation is done.  Note that
contrary to the previous discussion, the sum is over $W$ and not $|W|^2$.  In
this case, the normalization constant $N_P$ makes the method comparable to 
direct phase space tomography of one value of the Wigner function 
at one phase space point.  With this technique,
there is no additional gain in adding up 
components.

A more refined strategy uses amplitude amplification \cite{amplification}.
It is a generalization of Grover's algorithm \cite{grover}.  The
latter starts from an equal superposition of $N$ states, and in $\sqrt{N}$
operations brings the amplitude along one direction close to one.
Amplitude amplification increases the amplitude of a {\em whole subspace}.
If $P$ is a projector on this subspace, and $\hat{V}$ is the operator
taking $|0\rangle$ to a state having some projection on the desired
subspace, repeated iterations of 
$\hat{V}(I-2|0\rangle\langle0|)\hat{V}^{-1}(I-2P)$ on $\hat{V} |0\rangle$
will increase the projection.  Indeed, if one write $\hat{V} |0\rangle
=P\hat{V} |0\rangle +(I-P)\hat{V} |0\rangle$, the result of one iteration is to
rotate the state toward $P\hat{V} |0\rangle$ staying in the subspace 
spanned by $P\hat{V} |0\rangle$ and $(I-P)\hat{V} |0\rangle$. If $a=
|P\hat{V} |0\rangle|^2$, one can check that after one iteration the 
state is 
$(4 a^2-3)P\hat{V} |0\rangle +(4 a^2-1)(I-P)\hat{V} |0\rangle$,
with a component along $(I-P)\hat{V} |0\rangle$ decreased by $4a^2$.

If $\hat{V}$ is chosen to be $\tilde{U}_{Wigner}   \hat{U}^t $ (where
$\tilde{U}_{Wigner}$ builds the Wigner transform), and $P$
to be a projector on the space corresponding to a square of size
$N_D \times N_D$,
the process of amplitude amplification will increase the total 
probability in the square, keeping the relative amplitude inside
the square.  This acts like a ``microscope'', increasing the total probability
of one part of the Wigner function but keeping the relative details correct.
The total probability in a square of size
$N_D \times N_D$, following the results shown in Fig.3-6,
should be of the order $N_D^2/N^2$.  Amplitude amplification will
therefore need $N/N_D$ iterations to bring the probability 
inside the square close to one. Then according to
Fig 3-6 $N_D^2$ measurements  are needed to get all relative amplitudes 
with good precision.  Total number of quantum operations is therefore 
$O(tN_D N)$ (up to logarithmic factors).  
This should be compared to the number of classical
operations, $O(tN)$ for the evolution of the wave function, and
$O(N_D N)$ for computing the Wigner function 
(construction of the Wigner function on a square 
of size $N_D^2$ needs only $N_D$ Fourier transforms of $N$ dimensional
vectors).  Both computations are therefore comparable for low $K$.
When the scaling $N^{\alpha}$ ($\alpha<2$) 
for the IPR of the Wigner function is
verified, then $N_D^{\alpha}$ measurements are enough to get 
the Wigner function on a quantum computer, and a small gain of $N_D^{2-\alpha}$
is present for the quantum algorithm.  If the classical algorithm to 
compute the evolution of the map
were more complex, i.e. of order $O(tN^k)$ with $k >1$ (note the
difference with the previous case where $k >2$ was needed) , 
then in this 
case the quantum algorithm will be better by an additional
 factor $N^{k-1}$ if $N_D$
and $t$ are kept fixed.

Our last strategy uses the wavelet transform.
This transform \cite{Daub,meyer} is based on the wavelet basis,
which differs from the usual Fourier basis by the fact that each basis
vector is localized in position as well as momentum, with different scales.
The basis vectors are obtained by translations and dilations of an original
function and their properties enable to probe the different scales of
the data as well as localized features, both in space and frequency. 
Wavelet transforms 
are used ubiquitously on classical computers
for data treatment.  Algorithms for implementing
such transforms on quantum computers were developed in 
\cite{WT1,WT2,WT3,terraneo}, and were shown to be efficient, requiring
polynomial resources to treat an exponentially large vector.
Effects of imperfections on a dynamical system based on the wavelet transform
were investigated in \cite{terraneo}.  
In the present paper, we implemented the 4-coefficient Daubechies wavelet 
transform ($D^{(4)}$), the most commonly used in applications, and applied
it to the two-dimensional Wigner function (\ref{wignerimage}).

The results in Fig.\ref{IPR05}-\ref{IPR2} show that the 
IPR of the wavelet transform of the 
Wigner function scales as $N^{\beta}$, with $\beta$ decreasing from 
$\beta\approx 2$ to $\beta \approx 1.4$ when the chaos parameter $K$
is increased (for $K=0.5$, with low level of chaos, the wavelet transforms
yield a compression factor of order $10$, but no visible asymptotic gain).  
This means that getting the most important coefficients
in the wavelet basis needs only $N^{\beta}$ measurements.
The quantum algorithm for getting them 
needs only $O(tN^{\beta})$ operations.  On a classical machine, 
the slowest part is still the computation of the Wigner function,
which scales as $O(N^2)$.  Therefore at fixed $t$ the gain is
polynomial, of order $O(N^{2-\beta})$.  However, recovering the coefficients
of the original Wigner function needs to use a classical wavelet transform
which needs $O(N^2)$ operations.  
Still, the wavelet coefficients give information about the 
hierarchical structures in the wave function, so obtaining them with
a better efficiency gives 
some physical information about the system.  

Therefore, as concerns the quantum computation of the Wigner
function, it seems a modest polynomial gain can be obtained 
by different methods, especially in the parameter regime where the system is 
chaotic, the most efficient method being the measurement of the wavelet
transform of the distribution, although the interpretation of the results
is less transparent.

\section{IV. Measuring Husimi functions}

As already noted in Section II, 
the Wigner function is comparable to a classical phase space distribution, 
but can take negative values.  It is known that it becomes non-negative 
when coarse-grained over cells of size $\hbar$.  
One way to do this 
coarse-graining is to perform a convolution of the Wigner function with a
Gaussian, giving the Husimi distribution \cite{husimi}:
\begin{equation}
\label{husimifct}
\rho_H(\theta_0,n_0)=|\langle \phi_{(\theta_0,n_0)}|\psi\rangle|^2
\end{equation}
where 
$\phi_{(\theta_0,n_0)}(\theta,n)=
A\sum_{n} e^{-(n-n_0)^2/4a^2-i\theta_0n}|n\rangle$
is a Gaussian coherent state centered on $(\theta_0,n_0)$ with width $a$
($A$ is a normalization constant).  An interesting
quantum algorithm was proposed in 
\cite{saraceno} to compute this distribution,
based on phase space tomography.  It uses a relatively complicated
subroutine which builds an approximation of coherent states on a separate
register. This method is similar to the Wigner function computation 
through an ancilla qubit analyzed in the preceding section, and gives 
comparable results.

In \cite{frahm}, a very fast quantum algorithm was proposed to 
build a modified Husimi function, which is defined by:
\begin{equation}
\label{husimiklaus}
\rho_H^{(p)}(\theta_0,n_0)=|\langle \phi_{(\theta_0,n_0)}^{(p)}|\psi\rangle|^2
\end{equation}
where 
$\phi_{(\theta_0,n_0)}^{(p)}(\theta,n)=
(1/N^{1/4})\sum_{n=n_0}^{n_0+\sqrt{N}-1} e^{-i\theta_0n}|n\rangle$
is a modified coherent state centered on $(\theta_0,n_0)$.
The convolution is not made any more with a Gaussian function, but with
a box function of size $\sqrt{N}$ in momentum.  
This implies a
very good localization in momentum, but in contrast in the
angle representation the amplitude decreases only as a power law
since the Fourier transform of the box function is the function 
$\frac{\sin x}{x}$.

This transform can be evaluated quite efficiently on a quantum computer
without computing the Wigner function itself.  Indeed, as shown in 
\cite{frahm}, it can be computed by applying a QFT to the first half of 
the qubits.  This partial Fourier transform uses 
$\frac{n_q}{4}(\frac{n_q}{2}+1)$ quantum elementary operations to 
build from a wave function $|\psi \rangle$ with $N=2^{n_q}$ components
the state 
\begin{equation}
\label{husimiklaus2}
|\psi_H \rangle =\sum_{\theta,n} H(\theta,n) |\theta\rangle |n\rangle
\end{equation}
where $\theta$ and $n$
take only $\sqrt{N}$ values each and $|H(\theta,n)|^2$ is the modified
Husimi function
(\ref{husimiklaus}).  
Performing the same task on
a classical computer needs $O(N\log(N)^2)$ operations.  We will concentrate
on this method to compute Husimi functions, since it seems to be the most
simple and easy to implement, and gives a good picture of the wave 
function as can be seen in the implementations made in \cite{frahm,lee}.

\begin{figure}[h!]
\includegraphics[width=.49\linewidth]{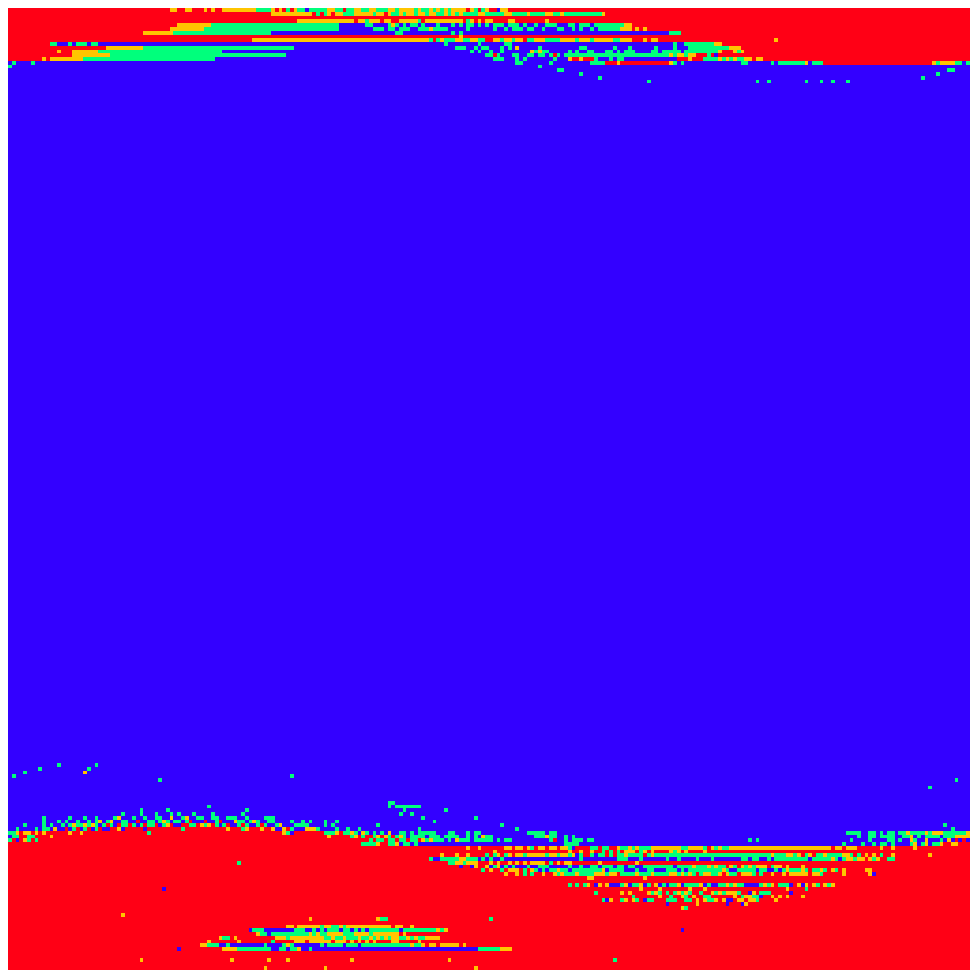}
\hfill
\includegraphics[width=.49\linewidth]{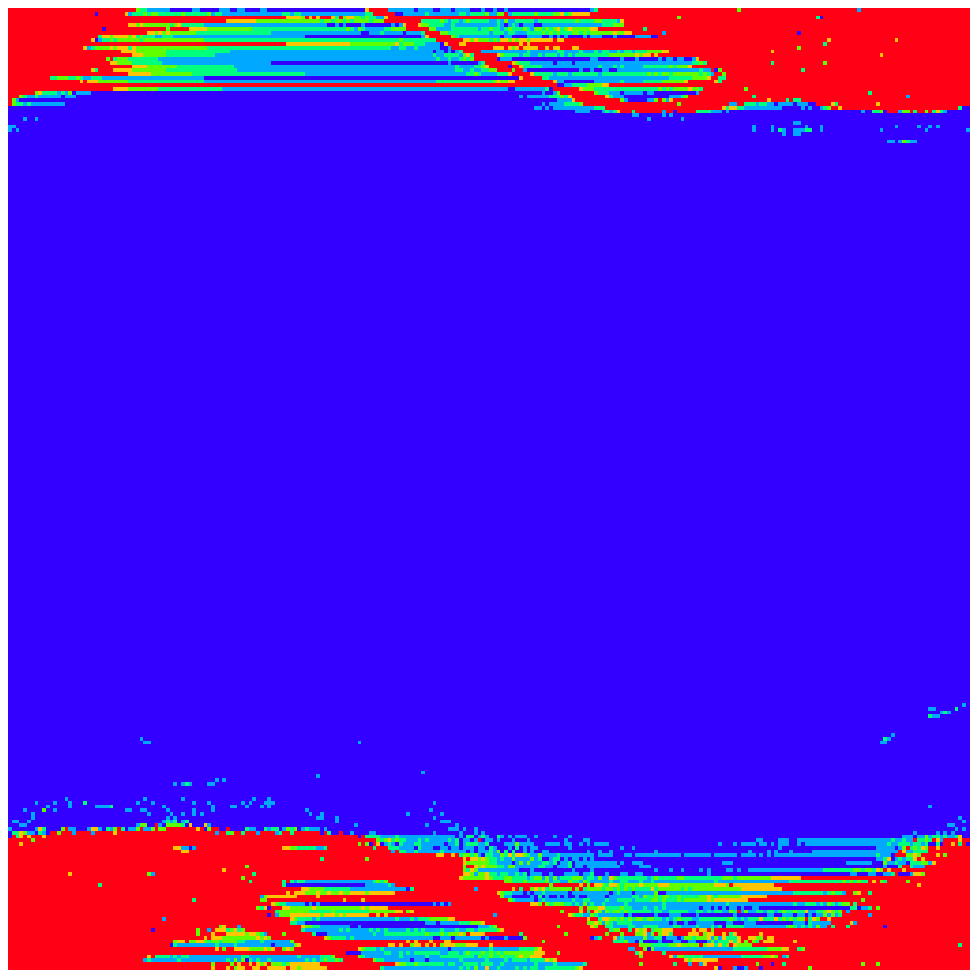}
\includegraphics[width=.49\linewidth]{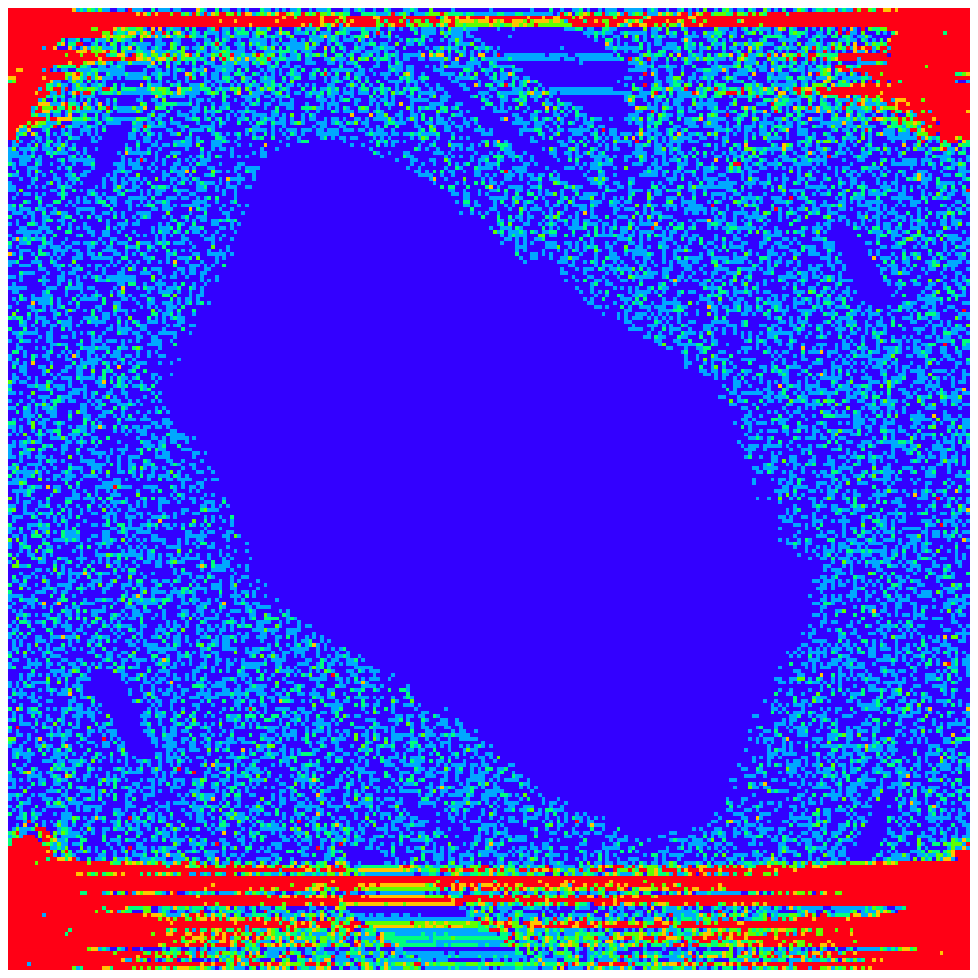}
\hfill
\includegraphics[width=.49\linewidth]{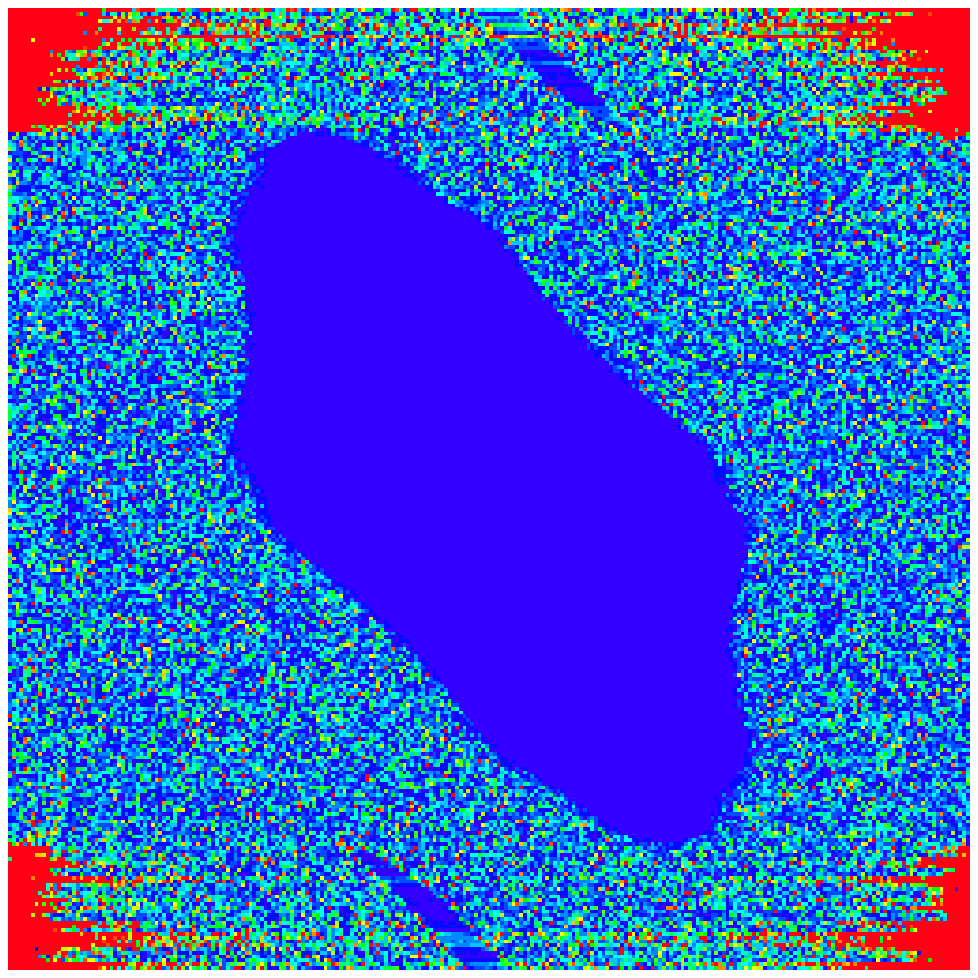}
\caption{(Color online)
Modified Husimi function (\ref{husimiklaus})
for the quantum kicked rotator with
$K=0.5$ (top left), $K=0.9$ (top right), $K=1.5$ (bottom left),
$K=2$ (bottom right). Here $T=2\pi/N$, where $N=2^{n_q}$, with $n_q=16$.
The function is plotted on a lattice of $\sqrt{N} \times
\sqrt{N}$ and each point is the average of $N$ points. Initial state 
is the same as in Fig.\ref{wignerf} (corresponding to the initial classical 
distribution in Fig.\ref{classical}), and the 
function is computed after $1000$ iterations of (\ref{qmap}).
Red (gray) is maximal value, blue (black) minimal value.
}
\label{klaus}
\end{figure}

In Fig.\ref{klaus}, we show the result of performing the evolution (\ref{qmap})
on a wave packet in $N$-dimensional Hilbert space
for four different values of $K$, and then applying
the partial Fourier transform.  The result is an array of 
$\sqrt{N} \times \sqrt{N}$ points, each point representing an average
over $\sim N$ neighboring values of the Wigner function.  The figure shows
that this transformation allows to obtain efficiently a positive 
phase space distribution which can be compared with the classical 
distributions, for example in Fig.\ref{classical}.

\begin{figure}[h]
\includegraphics[width=.8\linewidth]{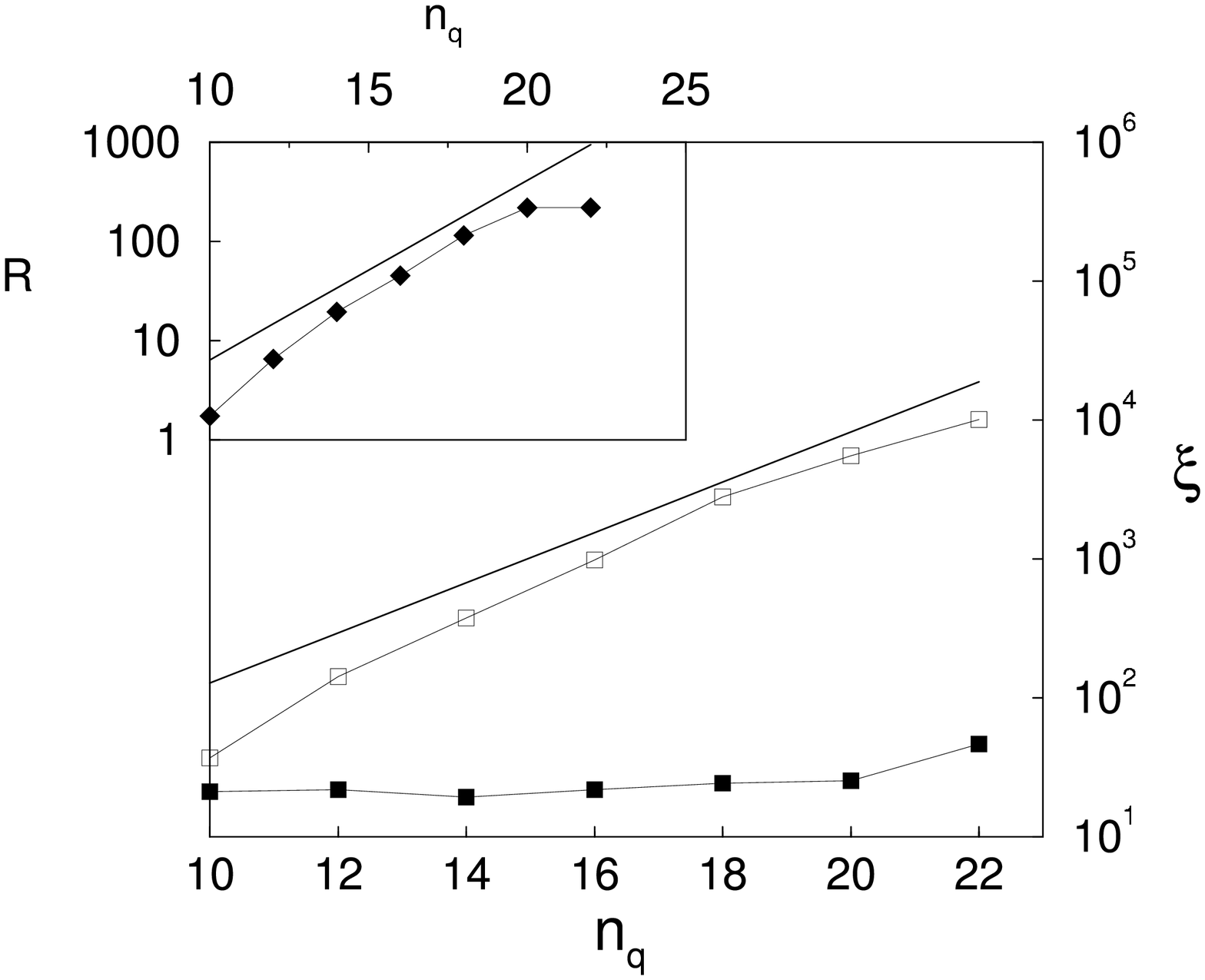}
\caption{Scaling for the IPR $\xi$ vs $n_q$ for
the function $H(\theta,n)$ in (\ref{husimiklaus2}),
 with
parameters $K=0.5$ and $T=2\pi/N$, $N=2^{n_q}$. 
Main plot: empty squares represent the IPR of  $H(\theta,n)$
function , the full squares represent the IPR of the wavelet transform 
of the modulus of $H(\theta,n)$.  The full line is $N^{0.6}$.
In the inset, the ratio $R$ between IPR of $H(\theta,n)$ and 
wavelet transform of $|H(\theta,n)|$ is plotted
 for different $n_q$, the full line is 
$N^{0.6}$. 
Number of iterations and initial state are the same as in Fig.\ref{klaus}.
}
\label{K05}
\end{figure}
\begin{figure}[h]
\includegraphics[width=.8\linewidth]{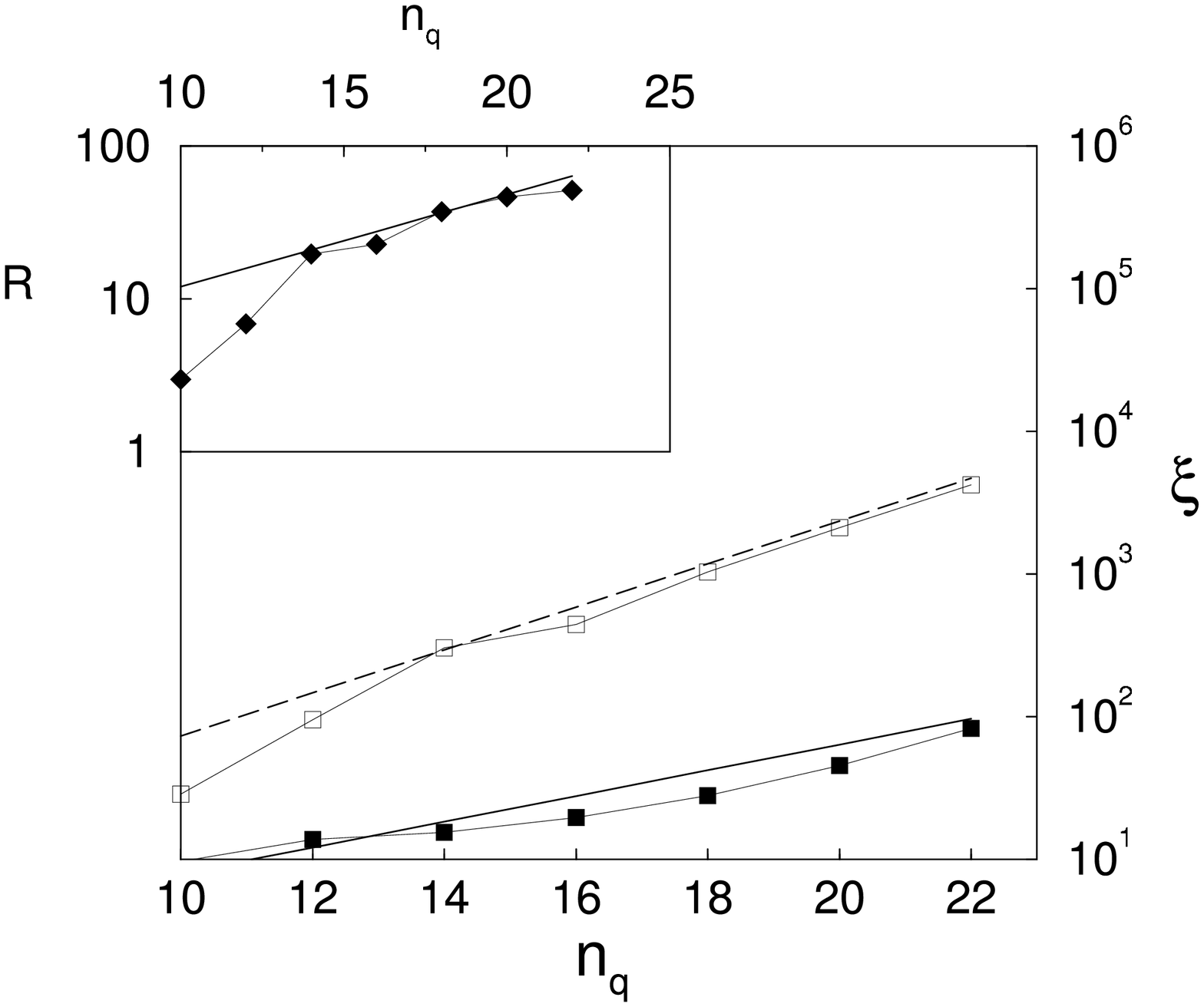}
\caption{Scaling for the IPR $\xi$ vs $n_q$ for
the function $H(\theta,n)$ in (\ref{husimiklaus2}), with
parameters $K=0.9$ and $T=2\pi/N$, $N=2^{n_q}$. 
Main plot: empty squares represent the IPR of $H(\theta,n)$, 
the full squares represent the IPR of the wavelet transform 
of the modulus of $H(\theta,n)$.  The full line is $N^{0.3}$, 
the dashed line is
$N^{0.5}$.
In the inset, the ratio $R$ between IPR of $H(\theta,n)$ and 
wavelet transform of $|H(\theta,n)|$ is plotted
 for different $n_q$, the full line is 
$N^{0.2}$. Number of iterations and initial state are the same 
as in Fig.\ref{klaus}.
}
\label{K09}
\end{figure}
\begin{figure}[h]
\includegraphics[width=.8\linewidth]{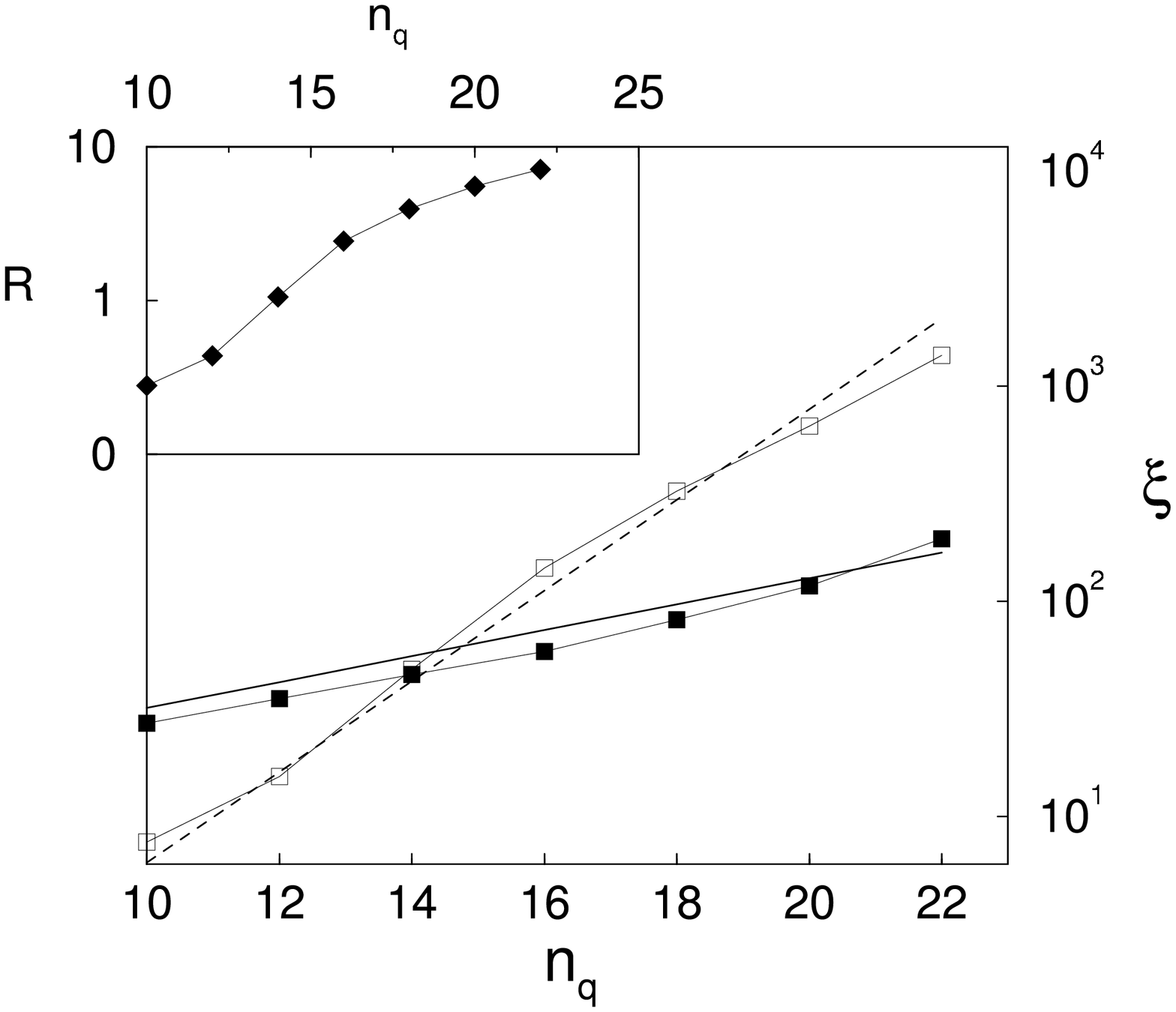}
\caption{Scaling for the IPR $\xi$ vs $n_q$ for
the function $H(\theta,n)$ in (\ref{husimiklaus2}), with
parameters $K=1.5$ and $T=2\pi/N$, $N=2^{n_q}$. 
Main plot: empty squares represent the IPR of $H(\theta,n)$, 
the full squares represent the IPR of the wavelet transform 
of the modulus of $H(\theta,n)$.
 The full line is $N^{0.2}$, the dashed line is
$N^{0.7}$.  In the inset, the ratio $R$ between IPR of $H(\theta,n)$
and 
wavelet transform of $|H(\theta,n)|$ is plotted
 for different $n_q$. Number of iterations and 
initial state are the same as in Fig.\ref{klaus}.
}
\label{figK15}
\end{figure}

\begin{figure}[h]
\includegraphics[width=.8\linewidth]{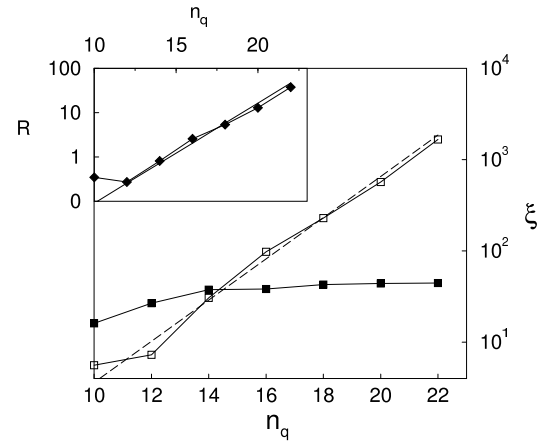}
\caption{Scaling for the IPR $\xi$ vs $n_q$ for
the function $H(\theta,n)$ in (\ref{husimiklaus2}), with
parameters $K=2$ and $T=2\pi/N$, $N=2^{n_q}$. 
Main plot: empty squares represent the IPR of $H(\theta,n)$, 
the full squares represent the IPR of the wavelet transform 
of the modulus of $H(\theta,n)$.  The dashed line is
$N^{0.7}$. In the inset, the ratio $R$ between IPR of $H(\theta,n)$
and 
wavelet transform of $|H(\theta,n)|$ is plotted
 for different $n_q$, the full line is $N^{0.7}$. Number of iterations and 
initial state are the same as in Fig.\ref{klaus}.
}
\label{K2}
\end{figure}

In Fig.\ref{K05}-\ref{K2} 
we show the IPR of the result of this transform
and of an additional wavelet transform of this function.
The data show that with this modified Husimi distribution the
compression of information is much better than in the case of the
Wigner function.

Indeed, in all four parameter regimes considered, the
IPR of the function scales as $N^{\gamma}$, with 
$ 0.5 \le \gamma \le 0.7$.
This means that the most important components of the modified
Husimi distribution can be measured with
$\sim N^{\gamma}$ quantum measurements.  Thus on a quantum computer
the whole process of evolving the wave function up to time $t$,
transforming it into the modified Husimi distribution and measuring it
needs $O(tN^{\gamma})$ operations.  On the contrary, a classical computer
will need $O(tN)$ operations for the evolution, and $O(N)$ for
the modified Husimi transform (up to logarithmic factors).
Thus for the system (\ref{qmap}), computation of
the modified Husimi transform is more
efficient on a quantum computer (including measurement) 
than on a classical one.  This gain would disappear if the transform
had IPR $\sim N$.

As in the preceding section, one can use coarse-grained
measurements in order to increase the probability.  
Again, this gives the
integrated probability inside the cells 
 in a number of measurements which scales 
with the number of cells, with the same drawbacks than in section III.

If we use amplitude amplification, the gain is even better.
Amplitude amplification will 
need $\sqrt{N/N_D}$ iterations to bring the probability 
inside a square of size $\sqrt{N_D}\times\sqrt{N_D}$ close to one.
Then according to Fig. 8-11 $N_D^{\gamma}$ measurements are needed,
with $0.5 \le \gamma \le 0.7$.
The total number of quantum operations is therefore 
$O(t \sqrt{N} N_D^{\gamma-1/2})$.
Classically, we still need $O(tN)$ operations for the evolution,
and $O(\sqrt{N}\sqrt{N_D})$ for the transform (up to logarithmic factors).
Thus for small $N_D$ a quadratic gain is achieved.  Interestingly enough,
this gain persists in the case where the IPR of the modified Husimi function 
is $\sim N$, even though the previous method then will not give any gain.
Since the IPR cannot be larger than $N$, this means that
with amplitude amplification the quantum computer is in general
at least quadratically faster at evaluating part of the modified
Husimi function
than any classical device.
 If the classical algorithm to 
compute the evolution of the map
were more complex, i.e. of order $O(tN^k)$ with $k>1$,
then in this 
case the quantum algorithm will be better by a
 factor $N^{k-1/2}$ if $N_D$
and $t$ are kept fixed, making the gain even larger, but still
polynomial.
 
We also analyzed the use of the wavelet transform to compress 
these data and minimize the number of measurements.  At this point a slight
complication appears.  In the previous section, individual
amplitudes of the wave function in (\ref{wignerimage}) were actual values
of the Wigner function, so performing a quantum wavelet transform of
(\ref{wignerimage}) was equivalent to a wavelet transform of the Wigner
function.  In the case at hand, the wave function of the quantum computer
is such that the {\it modulus square} of its components give the modified
Husimi distribution.  One can perform a quantum 
wavelet transform of this wave function, 
with real and imaginary parts for all coefficients,
which gives the wavelet coefficients of a complex-valued distribution whose
square is the modified Husimi distribution.  It is not clear how to
interpret the resulting coefficients, and anyway our data
have shown that this process does not decrease the IPR (data not shown),
thus making it an inefficient way of treating such data.
However, Figures 8-11 show that if one take the {\em modulus} of the
wave function, then the
IPR of the wavelet transform of this function is quite small, scaling as 
$O(N^{\delta})$, with $\delta \approx 0-0.2$.  So the modified
Husimi function
itself is well compressed by the wavelet transform.  It is the
phase of $H(\theta,n)$ in (\ref{husimiklaus2}) which, although
irrelevant for the Husimi functions, prevents compression by the wavelet
transform. 
To use efficiently the wavelet transform, we therefore need to
get rid of the phases, i.e. 
construct a wave function whose components are 
the {\em moduli} or
{\it moduli square} of the preceding wave functions.
  
Such a wave function can be prepared by
starting from two initial wave packets on two separate registers
$|\psi_0\rangle \otimes |\psi_0^* \rangle$, and as in the previous section
make them evolve independently to get 
$\hat{U}^t |\psi_0\rangle \otimes \hat{U}^{*t} |\psi_0^* \rangle$.
Then a partial Fourier transform is applied independently to both
registers, yielding 
$\sum H(\theta,n) H(\theta',n')^*
|\theta\rangle|\theta'\rangle |n\rangle |n'\rangle $.  
Then amplitude amplification should be used
to select the diagonal components, yielding 
$\sum |H(\theta,n)|^2 |\theta\rangle 
|n\rangle$. These components represented a probability $N/N^2=1/N$ of
the full original wave function, thus this process costs $O(t\sqrt{N})$
operations up to logarithmic factors.   This procedure gives us a final
wave function whose components are now the modified Husimi function itself,
without the irrelevant phases.  We can now apply the
quantum wavelet transform to this  wave function.
Afterward, measuring the main components
of the wave function should need only $O(N^{\delta})$ quantum measurements.
The cost of the total procedure is 
therefore $O(tN^{\delta+1/2})$ quantum operations, 
whereas classical computation will cost $O(tN)$ operations.
Obtaining the main wavelet components of this modified Husimi distribution
is therefore more efficient on a quantum computer than on a classical one,
albeit the gain is still polynomial.

\section{V. Standard images}

The investigations in the previous sections show that computation of
quantum phase space distributions can be more efficient on a quantum computer
than on a classical device.  An usually
 polynomial gain can be obtained for the whole
process of producing the distribution and measuring its values.  These
phase space distributions are in effect examples of two-dimensional
images.  It is interesting to explore these questions of efficiency
of image processing on a quantum computer in a more general setting.

\begin{figure}[h!]
\includegraphics[width=.49\linewidth]{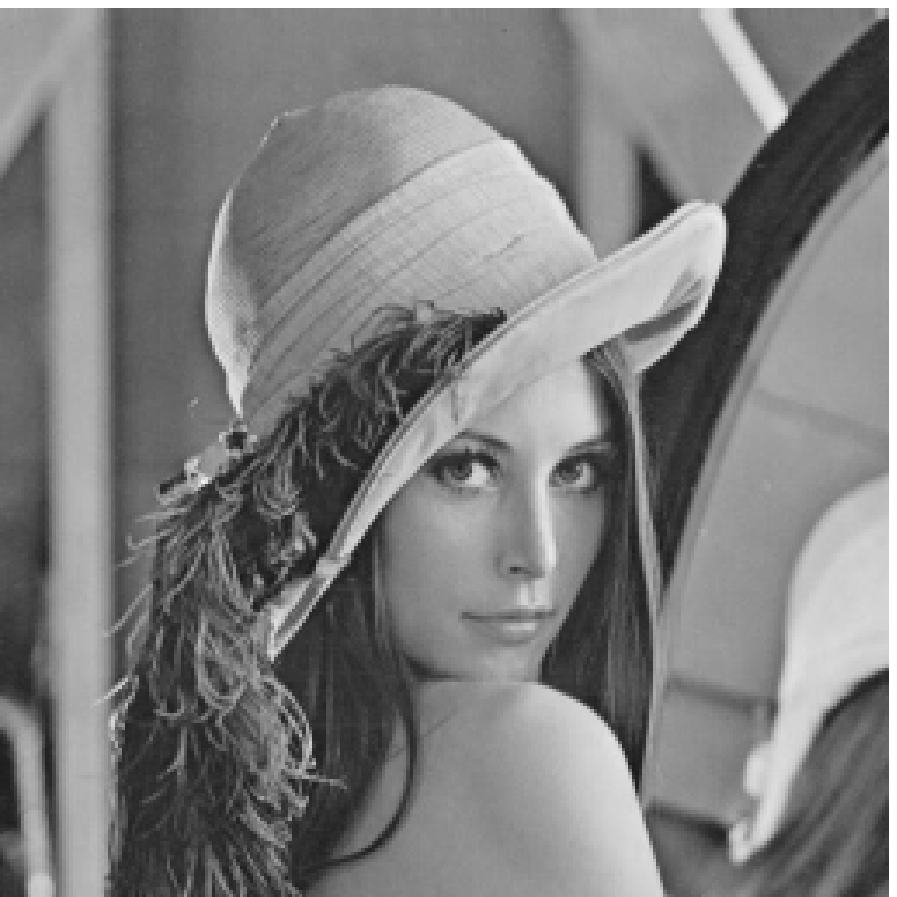}
\hfill
\includegraphics[width=.49\linewidth]{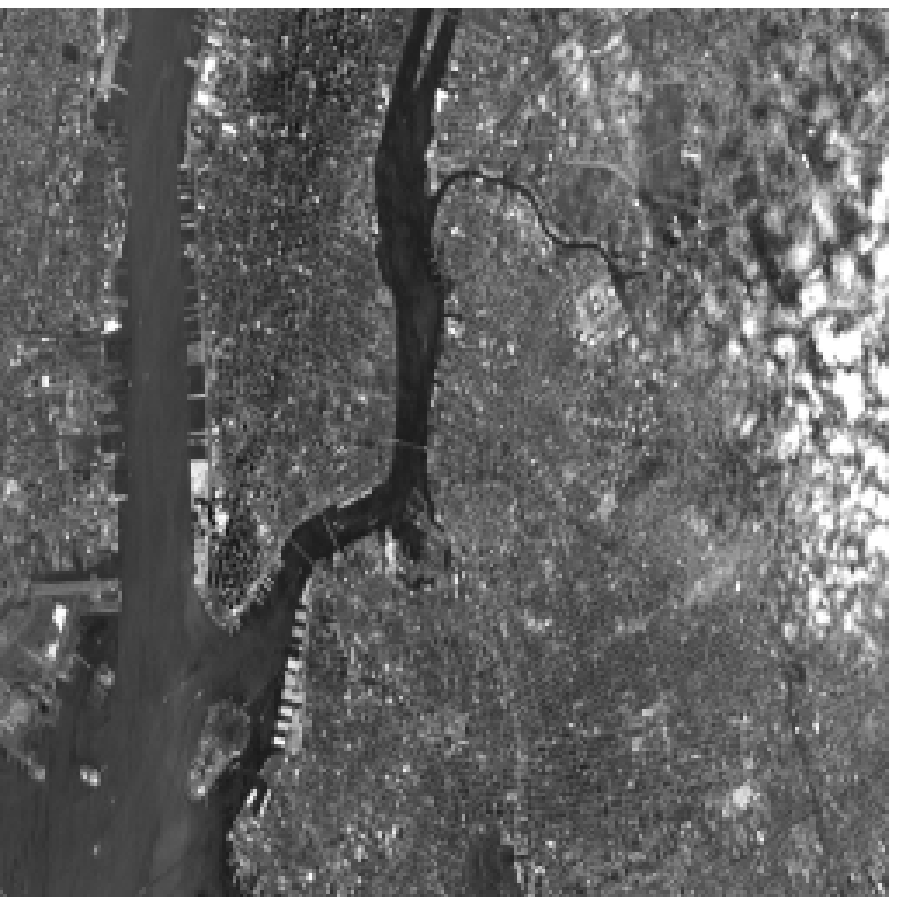}
\includegraphics[width=.49\linewidth]{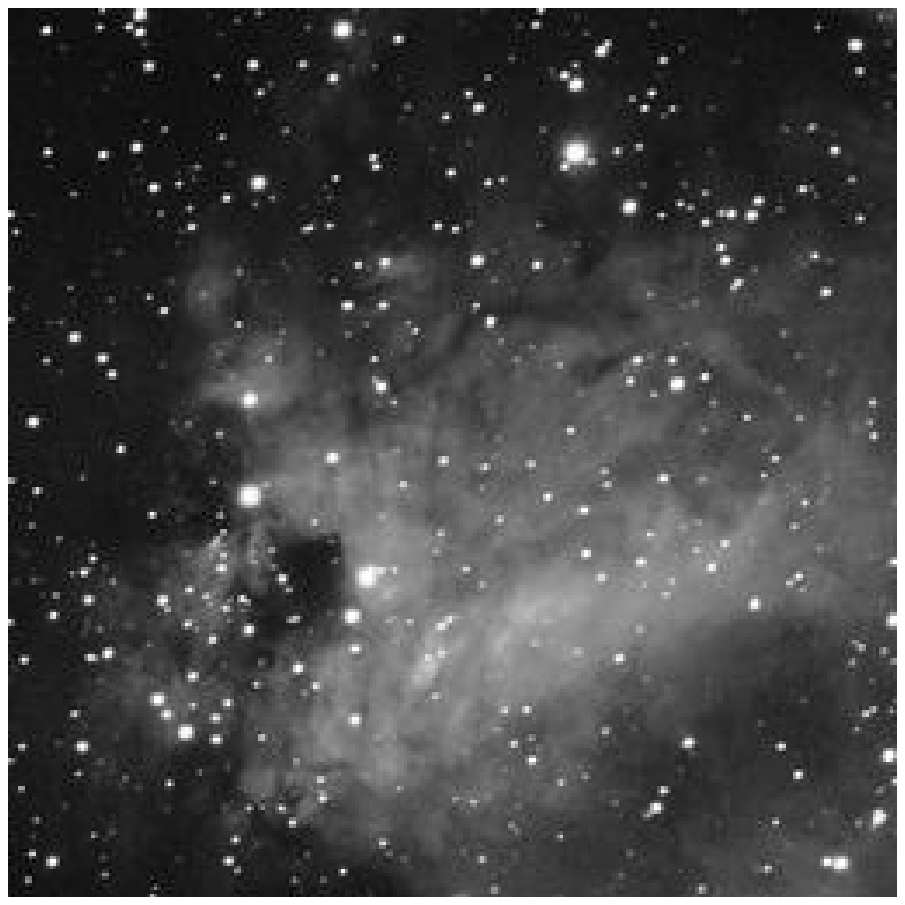}
\hfill
\includegraphics[width=.49\linewidth]{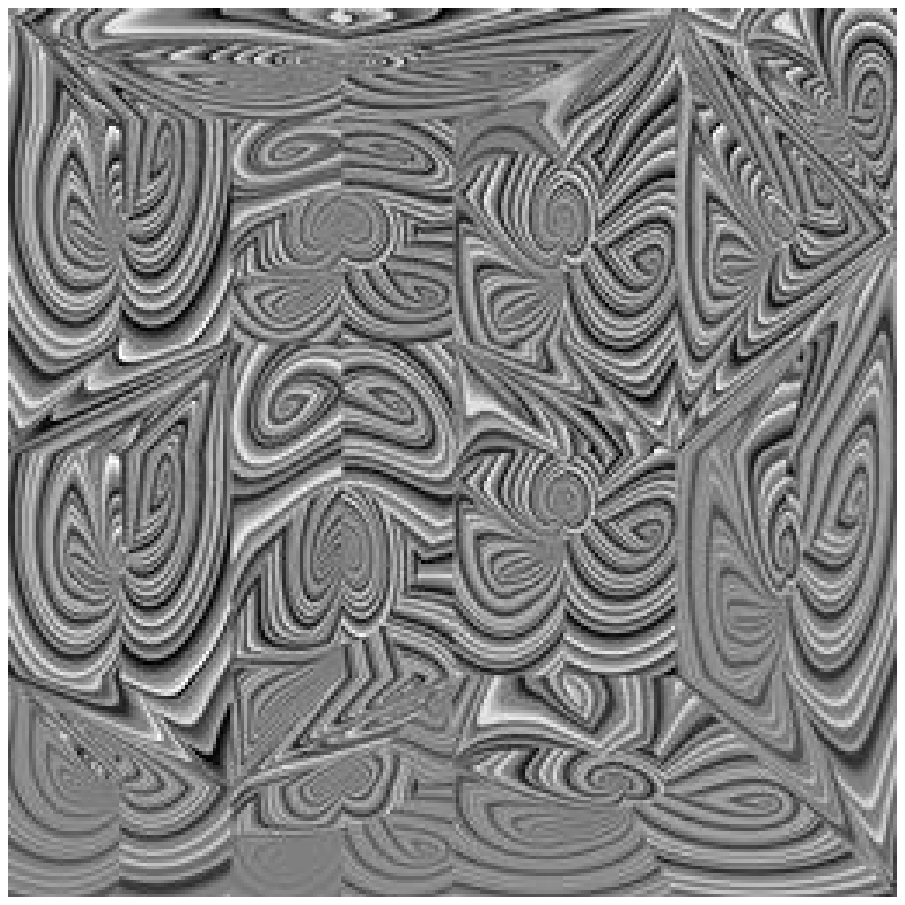}
\caption{
Images analyzed in this section. Top: girl image (left) and New York City
picture 
(right). Bottom: galaxy image taken from NASA website (left) and a fractal 
picture built on the purpose of studying image compression (right).
}
\label{standard}
\end{figure}

Fig.\ref{standard} shows four examples of classical images which we can
use as benchmark to test different strategies of processing them.
The top left image (the girl)
is a standard example used in the field of classical
image processing.  Top right is a aerial view of New York City,
bottom left an astronomical photograph, and bottom right an artificially built
picture with fractal-like structures.  They represent diverse types
of images that can be produced and processed for various purposes.
We will suppose in the following that these black and white pictures
are encoded on a quantum wave function in the form 
$\psi=\sum_{x,y} a_{xy} |x\rangle |y\rangle$ where $x,y$ are indexes of
$N^2$ pixels and $a_{xy}$ are the amplitudes on each pixel (positive number).
Of course, contrary to the previous examples, we do not know how to
produce in an efficient way such types of wave functions.  
We therefore concentrate on the problem of extracting information 
efficiently from such a wave function once it has been produced.

\begin{figure}[t!]
\includegraphics[width=.8\linewidth]{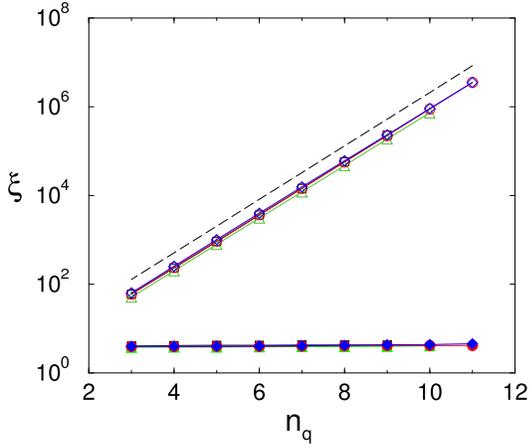}
\caption{(Color online) Scaling for the IPR $\xi$ vs $n_q$ for
the  images with different resolutions. 
Full symbols: $\xi$ after wavelet transform. Empty symbols:
 $\xi$ for the original images. 
with different resolution (from 32 $\times$  32 to 
2048 $\times$  2048). Squares refer to the
girl image, circles to the New York City image,
triangles to the galaxy image, and diamonds to the fractal image. 
The dashed line is the law $N^2$ with $N=2^{n_q}$. 
The original images are 8-bit gray scale images. They are encoded in the 
wave function from which the IPR is computed. 
}
\label{IPRstand}
\end{figure}

Fig.\ref{IPRstand} permits to analyze 
two of the strategies precedingly developed.  The IPR of the different images
are shown to scale like $N^2$, implying that direct measurement of all qubits
will need $O(N^2)$ measurements to get the most important components
(since these components scale also like $O(N^2)$).  
As in the case of the Wigner function, coarse-grained measurements
are possible, and require a number of measurements proportional to
the number of cells. This is more efficient, at the price
of losing information on scales smaller than the cell size.  

The use of amplitude amplification on a small part of the picture
(polynomial in $n_q$) enables to bring this part to a probability
close to $1$ in $O(N)$ Grover-like iteration.  So if one is interested
in details of the picture at a specific place predetermined, this
strategy is more efficient than the direct measurement.  Of course,
precise efficiency of the quantum process
compared to classical methods will depend on the relative
complexity of the classical and quantum image production, which 
probably varies with the problem.

The full symbols in Fig.\ref{IPRstand} give the IPR of the wavelet transform
of the image.  That is, the image is encoded in a quantum wave function
as previously, and a quantum wavelet transform is applied.  The
resulting wave function displays an IPR which grows slowly
with $n_q$.  Actually, data from
Fig.\ref{IPRstand} are compatible with a polynomial growth
with $n_q$ of the IPR.
  This would indicate that the wavelet transform is very
efficient in compressing information from standard images.  Obtaining
the main components of the wavelet transform would
demand polynomial number of 
measurements compared to an exponential one for
 the original image wave function.
This can transfer to an exponential gain in the full process if the
image can be encoded also in a polynomial number of operations in $n_q$.

\begin{figure}[t!]
\includegraphics[width=.8\linewidth]{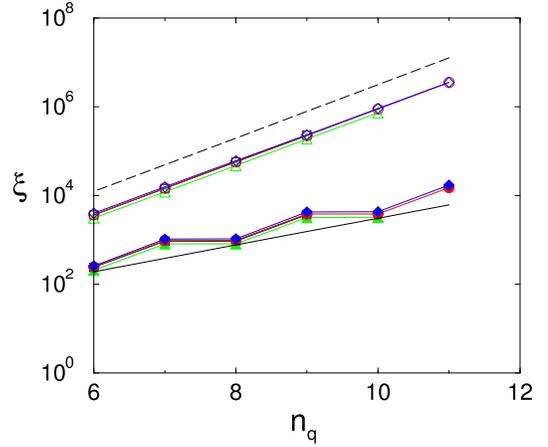}
\caption{(Color online) Scaling for the IPR $\xi$ vs $n_q$ for
the  images with different resolutions, the tiling method (see text) is used, 
with tiles of size $\sim \sqrt{N} \times \sqrt{N}$ 
Full symbols: $\xi$ after wavelet transform. Empty symbols:
 $\xi$ for the original images. 
with different resolution (from 32 $\times$  32 to 
2048 $\times$  2048). Squares refer to the
girl image, circles to the New York City image,
triangles to the galaxy image, and diamonds to the fractal image. 
The dashed line is the law $N^2$ with $N=2^{n_q}$, the full line
is the law  $N$.
The original images are 8-bit gray scale images. They are encoded in the 
wave function and the IPR is computed from the latter. 
}
\label{scaling}
\end{figure}

In Fig.\ref{scaling}, a different strategy is studied.  Namely, in analogy with
the MPEG standard for image compression, the image is decomposed into 
many tiles, and each tile is independently wavelet-transformed.
This procedure is tested in the case where tiles are of size 
$\sqrt{N} \times \sqrt{N}$.  Fig.\ref{scaling} shows that although
the final IPR grows more quickly with $n_q$ than in the case of
Fig.\ref{IPRstand}, the IPR seems asymptotically to be 
 smaller again than with the full image wave function.
Data from Fig.\ref{scaling} are compatible with an IPR growing like
$O(N)$, implying that the number of measurements is the square root 
of the one for the full wave function.  This suggests a
polynomial speed up with this method.  We note that a similar strategy for a 
quantum sound treatment was discussed in \cite{lee}.

\begin{figure}[t!]
\includegraphics[width=.8\linewidth]{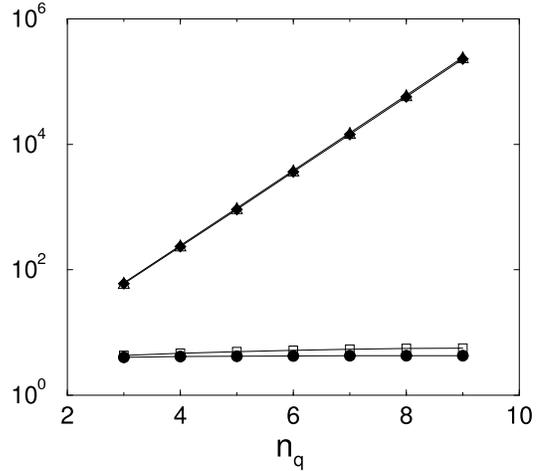}
\caption{
Comparison IPR / entropy for the girl image of Fig.\ref{standard}. 
Full symbols are for IPR, empty symbols for
$2^{S}$, where $S$ is the entropy. Squares and circles are for the wavelet
 transform, diamonds and triangles for the original image. Data for the three 
other images of Fig.\ref{standard} give the same result.
}
\label{entropy}
\end{figure}

Fig.\ref{entropy} enables to confirm the preceding results which use the IPR.
Indeed, an alternative quantity to quantify the spreading of a wave
function on a predetermined basis is the entropy.  For a  $N$-dimensional
wave function
$| \psi \rangle$ with projections on a basis $| \phi_j \rangle$ given by 
$W_j=|\langle \psi|\phi_j\rangle|^2$, the entropy is defined by 
$S=-\sum_j W_j \log_2 W_j$.  It takes values from $S=0$ ($\psi=\phi_j$
for some $j$) 
to $S=\log_2 N$ ($\psi=\frac{1}{\sqrt{N}}\sum_j |\phi_j\rangle$).
Both IPR and $2^S$ give an estimate of the number
of components of the wave function.
The data show that although both quantities are different, they show a similar
behavior with $n_q$ as do their wavelet transform, confirming that 
the preceding results are robust.

\begin{figure}[h!]
\includegraphics[width=.49\linewidth]{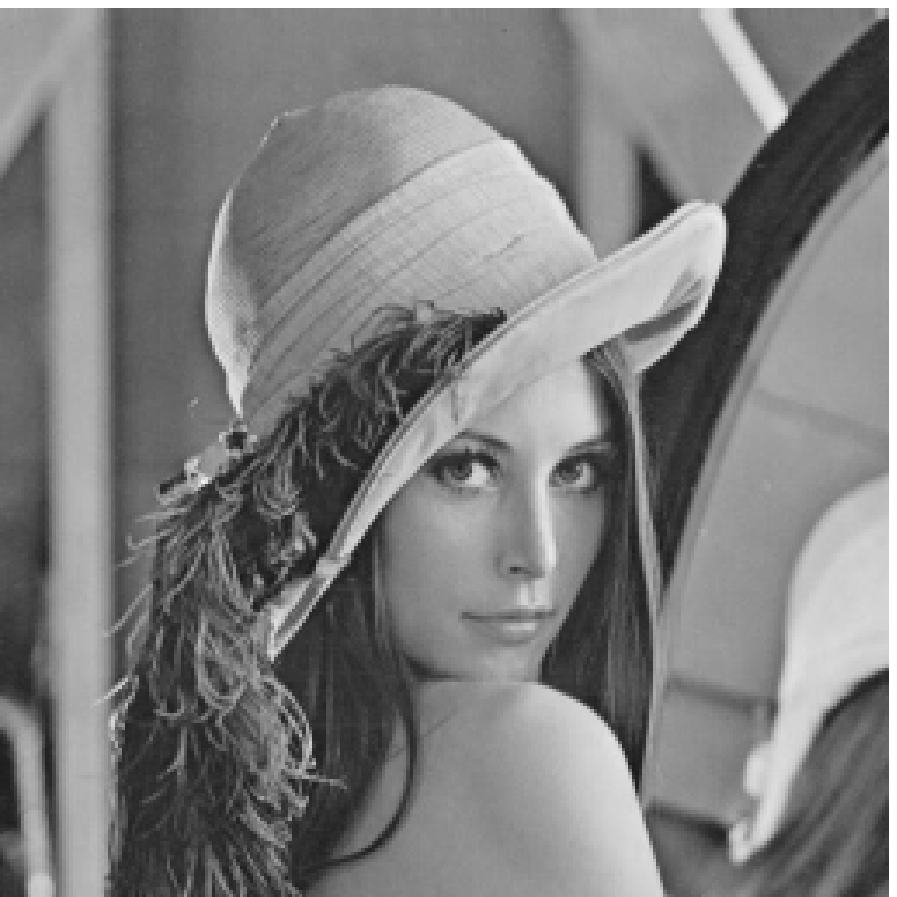}
\hfill
\includegraphics[width=.49\linewidth]{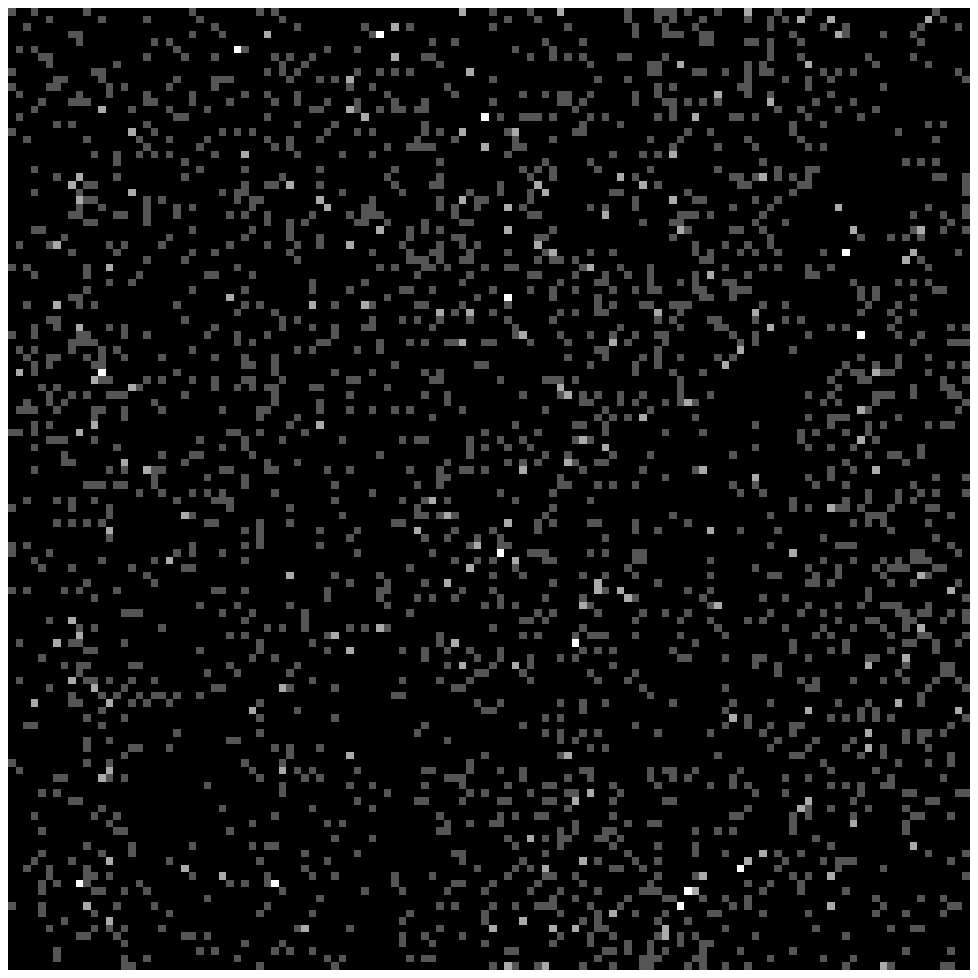}
\includegraphics[width=.49\linewidth]{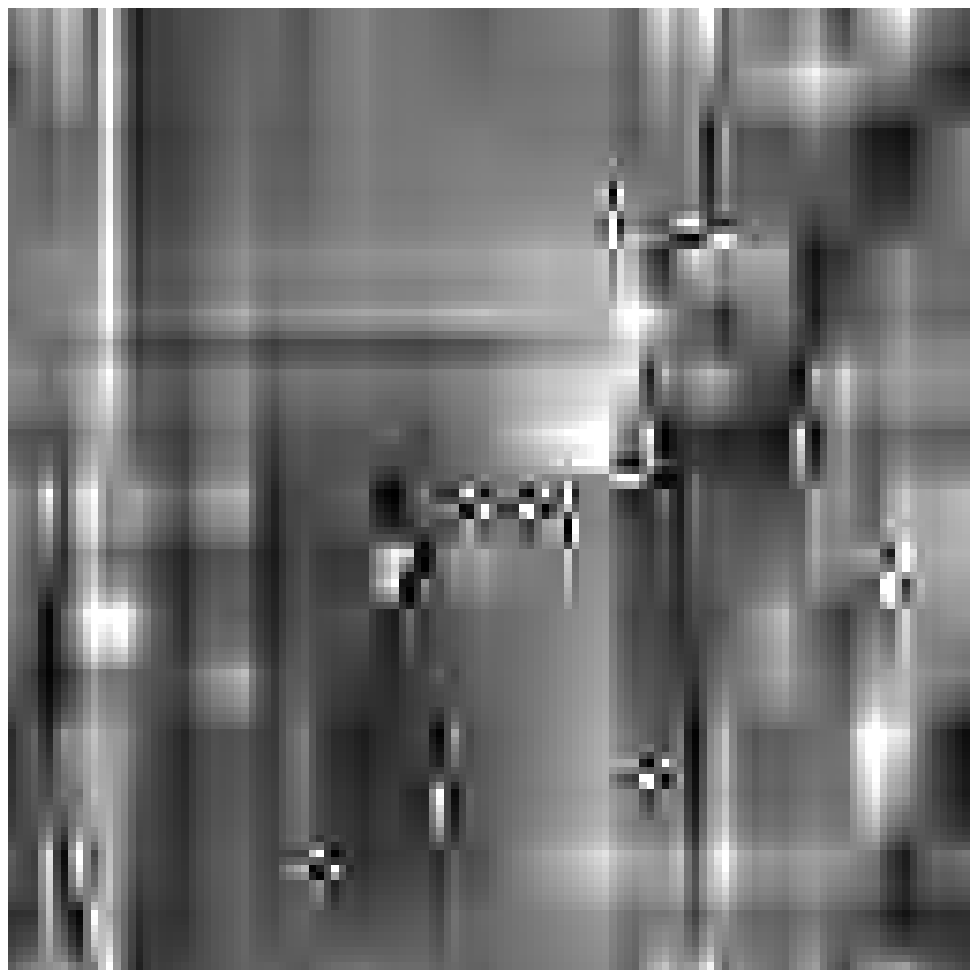}
\hfill
\includegraphics[width=.49\linewidth]{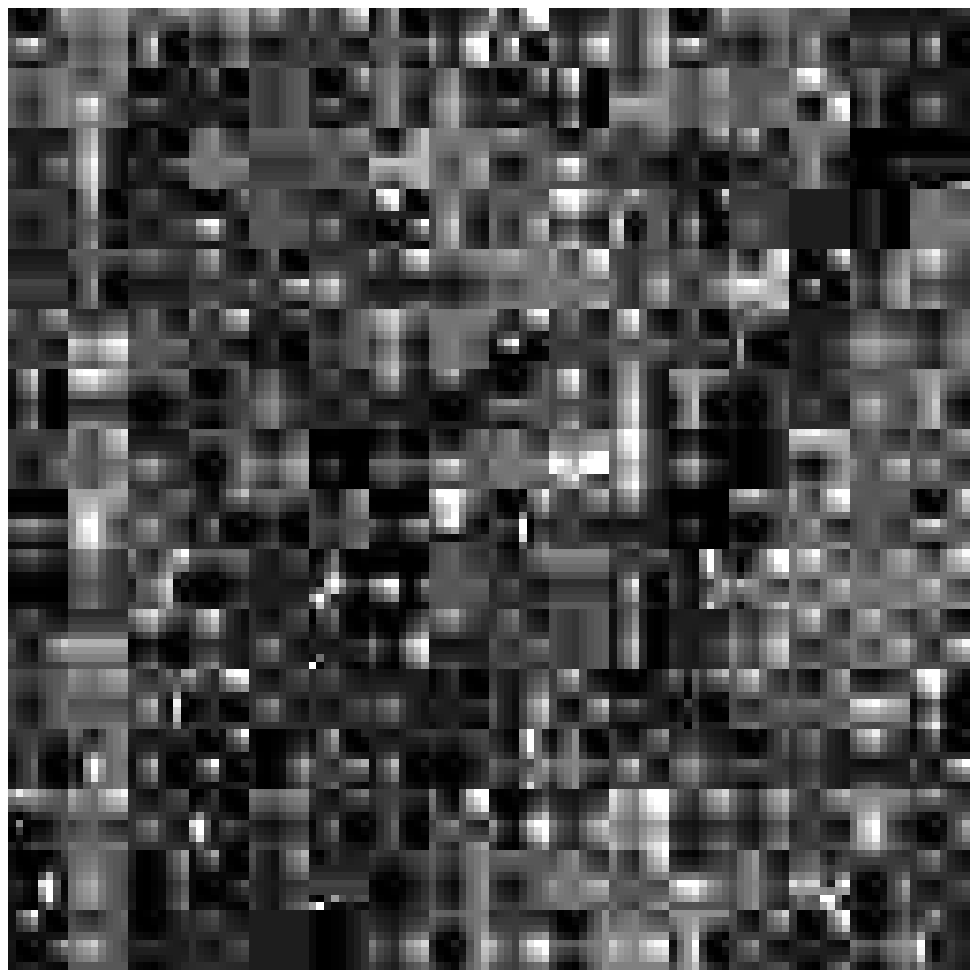}
\caption{
Image reconstruction from Monte Carlo sampling
and quantum wavelet sampling. Top: exact girl image (left) and 
sampled with $2500$ Monte Carlo points (right). 
Bottom: reconstruction after sampling with
 $2500$ measurements in the wavelet basis after full (left) and
tiled (right) wavelet transform. The images are $128 \times 128$ 
i.e. $\sim 16000$ points in total.
}
\label{figMC}
\end{figure}
\vspace{0.5cm}

The preceding discussion gives some numerical arguments 
suggesting that main components
of the wavelet transform can be obtained more efficiently than
the image itself.  This gives information on the patterns present in
the picture, and can be considered as an information in itself.  
It is also worth studying how much information about
this original image is present in these main components of the wavelet 
transform.  Fig.\ref{figMC} shows an attempt of reconstruction of one image
from these main components only.  The results displayed on this figure
show that although some features are distinguishable with this technique
(better than with the Monte-Carlo sampling), a lot of information
from the original figure has been lost.  It is possible that better results
are obtained for larger system sizes, but this 
regime cannot be reached by our classical numerical simulations.  Still,
even if the largest wavelet coefficients by themselves are not enough
to give a good approximation of the original image, they bring some
information about it
that can be obtained with a small number of measurements.

\section{VI. Conclusion}

In this paper, we have analyzed and numerically tested the quantum computation
of Wigner and Husimi distributions for quantum systems.  Two methods of
computation for the Wigner function, one original to this paper,
were considered.  We studied different strategies to extract information
from the wave function of the quantum computer, namely direct measurements,
coarse-grained measurements, amplitude amplification and measure
of wavelet-transformed wave function.  For the Wigner function, the 
largest (polynomial) gain is obtained through the use of the wavelet transform,
although other methods might yield a smaller gain in the chaotic regime.
For the Husimi distribution, the gain is much larger, although it is still
polynomial, and increases with the use of amplitude amplification
and wavelet transforms.  At last, the study of real images
show that the wavelet transform enables to compress information and therefore
to lower the number of measurements in the quantum case, 
although a lot of information is lost
in the process.

One of the authors (M.T.) acknowledges Benjamin L\'evi and Stefano
Gagliano for useful discussions about classical image treatment, 
and for helping him
in finding the high-resolution images in Fig.\ref{standard}.
We thank the IDRIS in Orsay and CalMiP in Toulouse for 
access to their supercomputers.
This work was supported 
by the EC  RTN contract HPRN-CT-2000-0156 and by the 
project EDIQIP of the IST-FET program of the EC.



\begin{thebibliography}{99}
\bibitem{nielsen} M.~A.~Nielsen and I.~L.~Chuang, {\it Quantum Computation
                   and Quantum Information}
                   (Cambridge University Press, Cambridge, 2000).
\bibitem{shor} P.~W.~Shor, in {\it Proceedings of the 
               35th Annual Symposium on the Foundations of
               Computer Science}, edited by S. Goldwasser
               (IEEE Computer Society, Los Alamitos, CA, 1994), p. 124.
\bibitem{grover} L.~K.~Grover, Phys. Rev. Lett. {\bf 79}, 325 (1997).
\bibitem{lloyd} S.~Lloyd, Science {\bf 273}, 1073 (1996); D.~S.~Abrams 
and S.~Lloyd, Phys. Rev. Lett. {\bf 79}, 2586 (1997).
\bibitem{schack} R.~Schack, Phys. Rev. {\bf A 57}, 1634 (1998).
\bibitem{GS}B.~Georgeot and D.~L.~Shepelyansky, Phys. Rev. Lett. {\bf 86}, 2890
  (2001). 
\bibitem{song} P.~H.~Song and D.~L.~Shepelyansky, Phys. Rev. Lett. {\bf 86},
               2162 (2001).
\bibitem{complex} G.~Benenti, G.~Casati, S.~Montangero and D.~L.~Shepelyansky, 
  Phys. Rev. Lett. {\bf 87}, 227901 (2001).
\bibitem{pomeransky}  A.~A.~Pomeransky and D.~L.~Shepelyansky,  
                      Phys. Rev. {\bf A 69}, 014302 (2004).
\bibitem{wigner} E.~Wigner Phys. Rev. {\bf 40}, 749 (1932); 
            M.~V.~Berry, Phil. Trans. Royal Soc. {\bf 287}, 237 (1977).
\bibitem{husimi} S.-J.~Chang and K.-J.~Shi, Phys. Rev. {\bf A 34}, 7 (1986).
\bibitem{Levi} B.~L\'evi, B.~Georgeot and D.~L.~Shepelyansky, 
            Phys. Rev. {\bf E 67}, 046220 (2003). 
\bibitem{harper} B.~L\'evi and B.~Georgeot,
                Phys. Rev. {\bf E 70}, 056218 (2004).
\bibitem{frahm} K.~M.~Frahm, R.~Fleckinger and D.~L.~Shepelyansky,
                Eur. Phys. J. {\bf D 29}, 139 (2004).
\bibitem{pazwigner} C.~Miquel, J.~P.~Paz, M.~Saraceno, E.~Knill, 
                     R.~Laflamme and
                    C.~Negrevergne, Nature {\bf 418}, 59 (2002).
\bibitem{saraceno} J.~P.~Paz, A.~J.~Roncaglia and M.~Saraceno,
                   Phys. Rev. {\bf A 69}, 032312 (2004).
\bibitem{amplification} G.~Brassard and P.~H{\o}yer, in
                   {\it Proceedings of Fifth Israeli Symposium 
                   on Theory of Computing and Systems} 
                    (IEEE Computer Society, Los Alamitos, CA, 1997)
                    pp. 12-23; 
                    G.~Brassard, P.~H{\o}yer, M.~Mosca and A.~Tapp, in
                    {\it Quantum Computation and Quantum Information:
                    A Millenium Volume}, edited by S.~J.~Lomonaco, Jr. and 
                    H.~E.~Brandt (AMS, Contemporary Mathematics Series Vol.
                    305, 2002).
\bibitem{Daub} I.~Daubechies, {\it Ten Lectures on Wavelets}, CBMS-NSF Series
 in Applied Mathematics (SIAM, Philadelphia, 1992). 
\bibitem{meyer} Y.~Meyer, {\it Wavelets: Algorithms and Applications} 
                (SIAM, Philadelphia, 1993).
\bibitem{WT1} P.~H{\o}yer, quant-ph/9702028 (1997).
\bibitem{WT2} A.~Fijaney and C.~Williams, Lecture Notes in Computer
              Science {\bf 1509}, 10 (Springer, 1998); quant-ph/9809004.
\bibitem{WT3} A.~Klappenecker, in {\it Wavelet Applications in Signal and
              Image Processing VII}, edited by M.~A.~Unser, A.~Aldroubi, 
              A.~F.~Laine, SPIE (1999), p. 703; quant-ph/9909014.
\bibitem{terraneo} M.~Terraneo and D.~L.~Shepelyansky, 
                   Phys. Rev. Lett. {\bf 90}, 257902 (2003). 
\bibitem{kolobov} M.~Kolobov, Rev. Mod. Phys. {\bf 71}, 1539 (1999).
\bibitem{negative} A.~I.~Lvovsky, H.~Hansen, T.~Aichele, O.~Benson, J.~Mlynek
                   and S.~Schiller, Phys. Rev. Lett. {\bf 87}, 050402 (2001).
\bibitem{qchaos} B.~V.~Chirikov, in
                {\it Les Houches Lecture Series},
               edited by  M.-J.~Giannoni, A.~Voros and J.~Zinn-Justin,
               (North-Holland, Amsterdam, 1991), Vol. 52.
\bibitem{lichtenberg} B.~V.~Chirikov, Phys. Rep. {\bf 52}, 263 (1979);
                A.~Lichtenberg and M.~Lieberman, 
               {\it Regular and Chaotic Dynamics}, (Springer, New York, 1992).
\bibitem{IEEE} G.~Casati, I.~Guarneri, and D.~L.~Shepelyansky, 
               IEEE Jour. of Quant. Elect. {\bf 24}, 1420 (1988);
               P.M. Koch and K.A.H. van Leeuwen, Phys. Rep.
               {\bf 255}, 289 (1995).
\bibitem{raizen} F.~L.~Moore, J.~C.~Robinson,
               C.~F.~Bharucha, B.~Sundaram and M.~G.~Raizen, 
               Phys. Rev. Lett. {\bf 75}, 4598 (1995).
\bibitem{loclength} G.~Benenti, G.~Casati, S.~Montangero and 
D.~L.~Shepelyansky, Phys. Rev.  {\bf A 67}, 052312 (2003).
\bibitem{discrete} C.~Miquel, J.~P.~Paz and M.~Saraceno
                Phys. Rev. {\bf A 65}, 062309 (2002).
\bibitem{lee} J.~W.~Lee, A.~D.~Chepelianskii and D.~L.~Shepelyansky,
              quant-ph/0309018 and in Proceedings of the SPIE conference 
 {\it Noise and information in nanoelectronics, sensors, and standards II}
edited by J.M.Smulko, Y.Blanter, M.I.Dykman, L.B.Kish, {\bf 5472}, 246 (2004).
\end{thebibliography}
\end{document}